\documentclass[a4paper,twocolumn,11pt]{quantumarticle}
\pdfoutput=1
\usepackage[utf8]{inputenc}
\usepackage[english]{babel}
\usepackage[T1]{fontenc}
\usepackage{amsmath}
\usepackage{hyperref}
\usepackage[numbers, sort&compress]{natbib}
\usepackage{amssymb} 
\usepackage{float}
\usepackage{array}
\usepackage{hhline}
\usepackage{enumitem}

\newcommand{\Bstrut}[1][14pt]{\rule[-#1]{0pt}{0pt}}
\newcolumntype{C}{>{$}c<{$\Bstrut}}
\newcolumntype{P}[1]{>{\centering\arraybackslash}p{#1}<{\Bstrut}}

\usepackage{tikz}
\usepackage{circuitikz}
\usepackage{lipsum}

\usetikzlibrary{patterns}
\tikzset{meter/.append style={draw, inner sep=10, rectangle, font=\vphantom{A}, minimum width=30, minimum height=10, line width=.8, path picture={\draw[black] ([shift={(.1,.3)}]path picture bounding box.south west) to[bend left=50] ([shift={(-.1,.3)}]path picture bounding box.south east);\draw[black,-{Latex[length= 6mm, width=1mm]}] ([shift={(0.05,.35)}]path picture bounding box.south) -- ([shift={(.5,.3)}]path picture bounding box.north);}}}
\usetikzlibrary{decorations.pathreplacing}

\begin{document}

\title{Efficient QKD in Non-Ideal Scenarios with User-Defined Output Length Requirements}
\author{A. Martín-Megino}
\affiliation{Universidad Carlos III de Madrid, Avda. de la Universidad 30, 28911 Legan\'es, Spain}
\email{martinmeginoa@gmail.com}
\affiliation{IMDEA Networks Institute, Av. Mar Mediterráneo, 22, 28918 Leganés, Madrid, Spain}
\author{B. López}
\email{blanca.lopez@networks.imdea.org}
\affiliation{Universidad Carlos III de Madrid, Avda. de la Universidad 30, 28911 Legan\'es, Spain}
\affiliation{IMDEA Networks Institute, Av. Mar Mediterráneo, 22, 28918 Leganés, Madrid, Spain}
\author{I. Vidal}
\email{ividal@it.uc3m.es}
\author{F. Valera}
\affiliation{Universidad Carlos III de Madrid, Avda. de la Universidad 30, 28911 Legan\'es, Spain}
\email{fvalera@it.uc3m.es}
\maketitle

\begin{abstract}
  Quantum Key Distribution (QKD) enables two parties to securely share encryption keys by leveraging the principles of quantum mechanics, offering protection against eavesdropping. In practical implementations, QKD systems often rely on a layered architecture where a key manager stores secret key material in a buffer and delivers it to higher communication layers as needed. However, this buffer can be depleted under high demand, requiring efficient replenishment strategies that minimize resource waste.
Given the importance of optimizing time and resources in quantum cryptography protocols, we introduce a variable-length adaptation of the BB84 protocol designed to meet user-defined output key length constraints in non-ideal scenarios. We present a method for dynamically configuring the protocol's initial parameters to generate secret keys of a desired length. To validate our approach, we developed simulation tools to model general QKD networks and discrete-variable protocols. 
These tools were used to implement and evaluate our strategies, which were developed within the BB84 framework but can be extended to other QKD protocols under reasonable assumptions. The results highlight their usefulness in optimizing quantum resource usage and supporting key management, contributing to the long-term goal of scaling and strengthening secure quantum networks.
\end{abstract}

\section{Introduction}\label{Introduction}
Quantum computers could potentially break modern cryptography. This has become a significant concern in recent years and has emerged as a topic of discussion across several fields, driven by rapid advances in quantum computing over the past decade. However, rather than simply highlighting the risk, it is important to understand why this constitutes such a significant threat.

The new computational paradigm introduced by quantum computing leverages quantum information theory to solve certain mathematical problems with unprecedented efficiency, significantly surpassing even the most advanced classical algorithms. A relevant example is Shor’s algorithm \cite{shor1999polynomial}, a quantum circuit that enables polynomial-time integer factorization and discrete logarithm calculation, tasks that classical methods can only perform in exponential time. The mere possibility of implementing this algorithm on a sufficiently powerful quantum computer poses a serious threat to asymmetric cryptography, which relies on the infeasibility of efficiently solving these mathematical problems. 
Experts have long suggested that the year 2030 could mark the transition from experimental prototypes to practically useful quantum computers, and recent breakthroughs make this timeline appear more realistic than previously expected \cite{acharya2024quantum, bluvstein2024logical}.  
Therefore, the potential of Shor’s algorithm to break cryptographic standards such as RSA \cite{rivest1978method}, ECC \cite{miller1985use, koblitz1987elliptic}, and Diffie-Hellman \cite{diffie1976new} has already turned a theoretical threat into an urgent security concern \cite{gidney2025factor}.
 In addition to Shor’s, another relevant example is Grover’s algorithm \cite{grover1996fast}. Though less severe in its impact, it weakens symmetric encryption protocols like AES \cite{daemen1999aes}.


As companies race to develop powerful quantum computers, digital data becomes an increasingly valuable asset, and advances in number theory continue to challenge the assumptions underlying classical cryptography, an urgent response is needed to safeguard future communications. In this regard, two main approaches are being actively explored. 
The first is Post-Quantum Cryptography (PQC), which focuses on developing classical cryptographic algorithms that can withstand attacks from quantum computers. As part of its standardization effort, NIST released in August 2024 a final list of three PQC tools: ML-KEM \cite{FIPS203}, intended as the standard for general encryption and based on the CRYSTALS-Kyber algorithm \cite{bos2018crystals}; ML-DSA \cite{FIPS204}, the primary standard for digital signatures, based on the CRYSTALS-Dilithium algorithm \cite{ducas2018crystals}; and SLH-DSA \cite{FIPS205}, also designed for digital signatures and derived from the Sphincs+ algorithm \cite{bernstein2019sphincs+}. These algorithms are expected to serve as the backbone of secure communication in the post-quantum era, and NIST is already encouraging system administrators to start the transition to these new standards. However, their long-term security remains uncertain, as they are relatively new and could be compromised by future advances in the field.

The second approach is Quantum Key Distribution (QKD), a set of methods that utilize the principles of quantum mechanics to establish symmetric secret keys. Unlike classical techniques, QKD offers information-theoretic security, guaranteed by the fundamental laws of quantum mechanics rather than relying on the assumed difficulty of solving specific mathematical problems \cite{lo1999unconditional}. This unprecedented level of security is achieved by encoding key information in quantum bits (qubits) and exploiting quantum properties such as superposition and entanglement. Over the past decades, QKD has progressed from theoretical research \cite{BEN84, ekert1991quantum, grosshans2002continuous, scarani2004quantum, lo2012measurement} to experimental demonstrations \cite{bennett1989experimental, rosenberg2007long, zhang2020long} and initial commercial deployments, including the development of QKD networks \cite{elliott2005current, peev2009secoqc, sasaki2011field, chen2021integrated}, where multiple nodes can establish shared secret keys by executing QKD protocols.

\begin{figure}[h]
    \centering
    \includegraphics[width=\columnwidth]{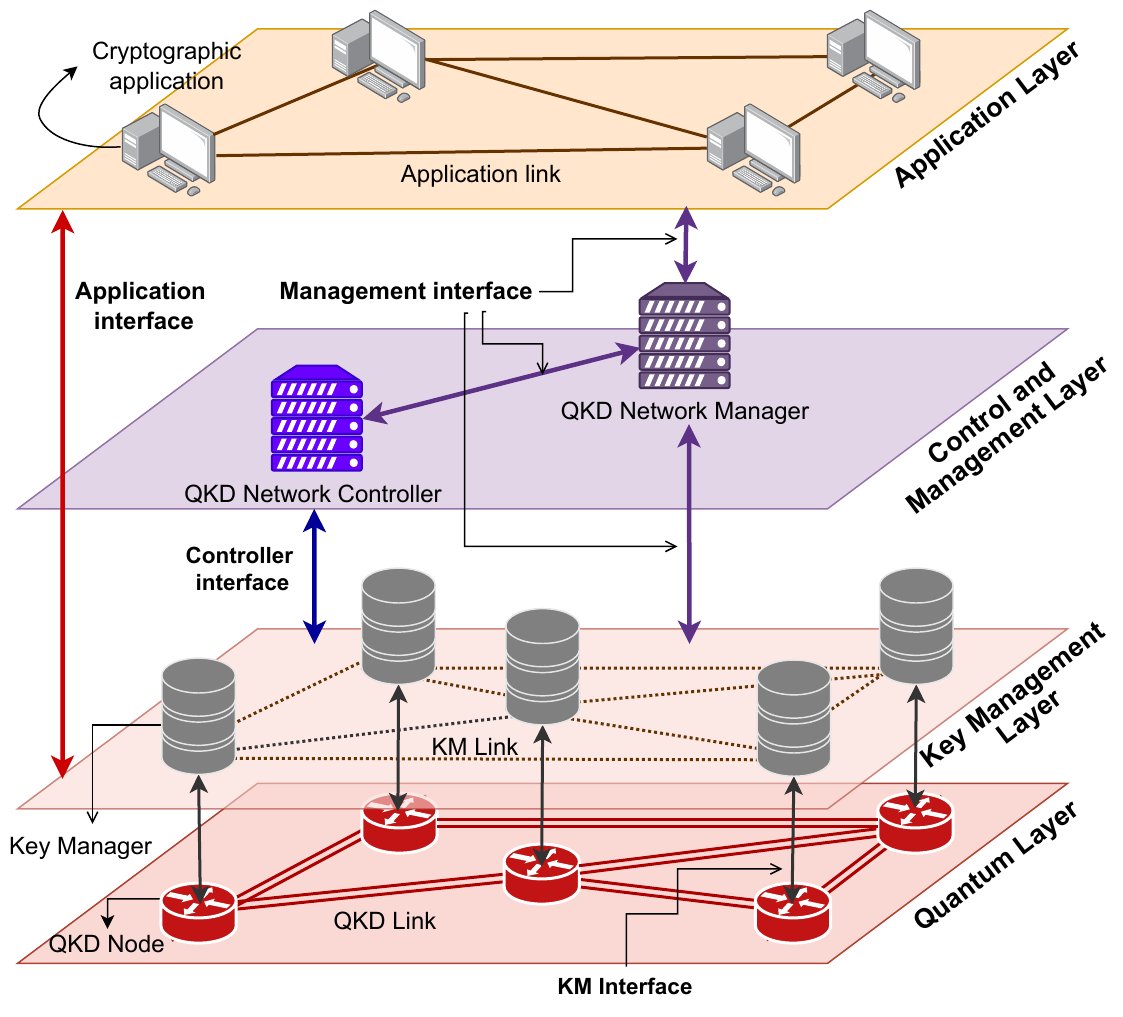}
    \caption{General architecture of QKD networks.}
    \label{fig:QKDNs}
\end{figure}

A QKD network (QKDN) consists of interconnected nodes linked via optical fiber or free-space channels to execute QKD protocols. QKDNs are considered a key step towards the realization of the quantum Internet \cite{kimble2008quantum, wehner2018quantum}, a future global network of quantum information processors designed to interconnect next-generation quantum computers. Conceived as a complement to the classical Internet, the quantum Internet will enable end-to-end QKD and support novel network functionalities beyond those of today’s communication systems \cite{giovannetti2004quantum, broadbent2009universal, jozsa2000quantum}. 
To take full advantage of the benefits offered by QKD protocols while ensuring efficient key management, storage, and integration with classical infrastructure, it is essential to design a layered architecture tailored to this emerging technology (see Fig.~\ref{fig:QKDNs}). Various approaches have been proposed as the design of QKD network architectures remains an open topic, and a dedicated ETSI report is expected by the end of 2025 \cite{ETSI_GS_QKD_017_V1.1.1}. Nonetheless, there is broad consensus on including QKDN management and control layers between the user or application layer \cite{ITU-T_Y3800_2019, ITU-T_Y3801_2020} and the physical infrastructure. These two intermediate layers coordinate the dynamic operation and overall supervision of the network by handling session control, key relay routing, link configuration, performance monitoring, and enforcement of security and quality policies \cite{ITU-T_Y3804_2020}. 

Below the QKDN control layer lies the key management layer \cite{ITU-T_Y3803_2020, ETSI_GS_QKD_014_V1.1.1,ETSI_GS_QKD_004_V2.1.1}. Each key manager (KM) is located within a corresponding QKD node and is connected to other KMs via classical KM links. It is responsible for securely storing the classical bit strings generated by the QKD modules in the physical or quantum layer, and managing IT-secure key relay between nodes by requesting an appropriate relay route from the QKDN controller. The KM also oversees the key lifecycle and handles requests from cryptographic applications by retrieving key material from storage, formatting and synchronizing it, and authenticating it via classical communication. The stored key material is periodically updated and, to the best of our knowledge, continuously replenished to ensure it can meet application demand at all times. 
This is not desirable from a security perspective, as it involves retaining cryptographic material in memory longer than necessary. This is achieved by repeatedly running the protocol in fixed-size blocks, without considering the key manager's actual situation and needs. Generating keys closer to the moment they are needed is, in principle, a more secure strategy.

The need to efficiently manage time and quantum resources while meeting varying key demands leads us to reassess the interface between the QKD module and the KM. In our approach, instead of operating continuously to completely fill the secret key buffer, the QKD module executes the protocol only when triggered by the KM. 
The KM also determines the amount of key material needed for each request based on demand or other requirements, and configures relevant protocol parameters such as the number of exchanged quantum signals to generate keys of the desired length. 
We strongly believe that enhancing the flexibility and efficiency of the KM interface is crucial, since it is the bridge between quantum communication hardware and the supporting classical infrastructure. 

In this work, we tested our proposed strategies within the BB84 protocol framework \cite{BEN84}, the first QKD protocol ever introduced and still one of the most widely used today. 
As a first step, we developed a software environment to simulate non-ideal QKD network configurations and the execution of discrete-variable QKD protocols. This environment was built using \texttt{NetSquid}~\cite{coopmans2021netsquid} as the underlying simulation framework.
The theoretical background for this part of the study is presented in Sect.~\ref{2}, where we detail the non-idealities considered in our model and highlight several non-trivial aspects of the simulation setup.
In Sect.~\ref{3}, we introduce our variable-length adaptation of the BB84 protocol. This section outlines all stages of the protocol, defines important quantities, and clearly distinguishes between the quantum and post-processing phases.
We then shift focus in Sect.~\ref{4} to formalize the problem to be solved within this scenario. Assuming a momentary key demand of $m_F$ key bits, we derive the theoretical framework needed to determine the minimum number of single-photon pulses Alice (the sender) must send to Bob (the receiver) to ensure that the resulting final key $K_F$ satisfies $m = \mathrm{length}(K_F) \geq m_F$ with a predefined confidence level and while preserving information-theoretic security.
Simulation results supporting this analysis are presented in Sect.~\ref{5}, and we conclude with a summary and discussion of our findings in Sect.~\ref{6}.

\section{Modeling and Simulation of Non-Ideal QKD Protocols}\label{2}

Current QKD technology supports secure communication over direct links spanning tens to hundreds of kilometers, despite facing several practical challenges. First, the no-cloning theorem prohibits the duplication of arbitrary quantum states, disallowing classical signal amplification techniques. This fundamental limitation, combined with photon loss and decoherence, restricts the maximum communication distance achievable by direct QKD links \cite{pirandola2017fundamental}. In addition to these channel non-idealities, quantum hardware at both the sender node (Alice) and the receiver node (Bob) is also conditioned by emission and detection efficiency, state preparation quality, dark count rates, detector dead time, and detector delay, among others \cite{ETSI_GS_QKD_003_V2.1.1}. Furthermore, in this work we focus exclusively on protocols where information is encoded in discrete quantum observables, such as photon polarization. These protocols are ideally defined using perfect single-photon sources, but such sources are not available in practice. Real-world implementations typically use highly attenuated laser pulses, which approximate single-photon emission, though the number of photons per pulse follows a Poisson distribution. Although the average photon number $\mu$ is set below 1, multiphoton events can still occur, making the system vulnerable to photon number splitting (PNS) attacks \cite{lutkenhaus2002quantum}. This posed a significant security concern for the implementation of discrete-variable QKD protocols until the decoy-state solution was proposed \cite{lo2005decoy, ma2005practical, wang2005beating}.  For simplicity, we do not consider multiphoton pulses in our work. We will only account for the probability of sending 0 photons (the vacuum state) when defining the emission efficiency $\eta_e$. While this simplification does not invalidate our analysis of QKD performance, it is important to acknowledge this limitation when simulating a realistic scenario.

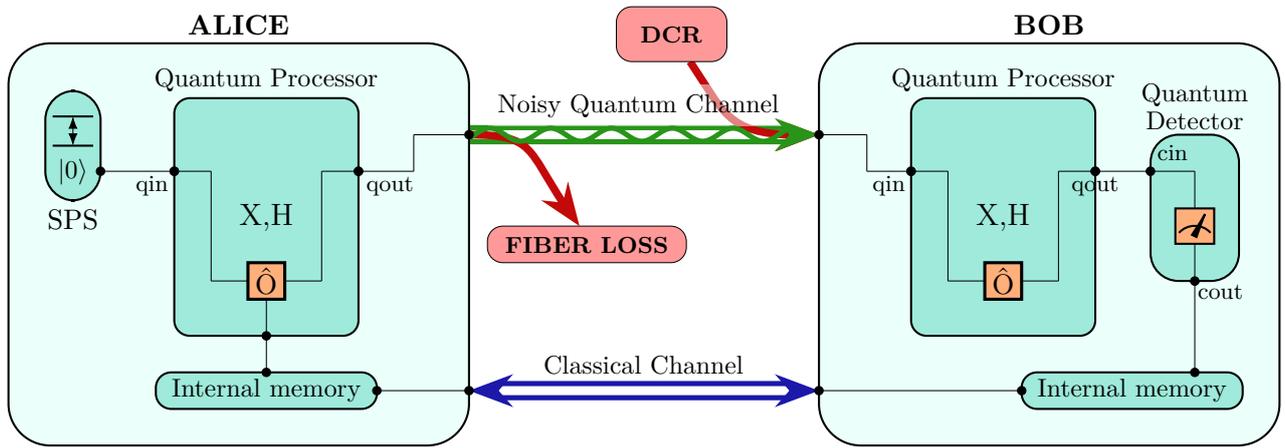
\begin{figure*}[t!]%
  \centering 
\resizebox{\linewidth}{!}{
\begin{circuitikz}
    \tikzstyle{every node}=[font=\normalsize]
        
        \draw [color={rgb,255:red,27; green,24; blue,170}, line width=2pt, <->, >=Stealth,double distance=3.5pt] (0.75,20.75) -- (5.5,20.75);
        \draw [color={rgb,255:red,40; green,148; blue,25}, line width=2pt, ->, >=Stealth,double distance=3.5pt] (0.5,24.25) -- (5.5,24.25);
        \draw [ color={rgb,255:red,195; green,9; blue,9}, line width=3pt, ->, >=Stealth] (0.75,24.25) .. controls (1.75,24.25) and (1.5,24) .. (2.25,23) ; 
        \draw [ color={rgb,255:red,195; green,9; blue,9}, line width=3pt, ->, >=Stealth] (3.75,25.25) .. controls (4.25,24.5) and (4.25,24.2) .. (5.5,24.25) ;
        \node [font=\small, fill={rgb,255:red,255; green,255; blue,255}, fill opacity = 0.5, text opacity = 1] at (3.05,24.65) {Noisy Quantum Channel};
        \draw [color={rgb,255:red,40; green,148; blue,25}, line width=2pt, ->, >=Stealth,double distance=3.5pt] (5.1,24.25) -- (5.5,24.25);
        \draw[domain=0.5:5.15, samples=100, color={rgb,255:red,40; green,148; blue,25}, line width=2pt] plot (\x,{0.09*sin(7.61*\x r  ) +24.25});

        \draw [fill={rgb,255:red,235; green,255; blue,251}, rounded corners=16.2, line width = 0.8] (-5.5,25.5) rectangle (0.75,20);
        \draw [fill={rgb,255:red,235; green,255; blue,251}, rounded corners=16.2, line width = 0.8] (5.5,25.5) rectangle (11.75,20);
        \draw [fill={rgb,255:red,159; green,234; blue,219}, rounded corners=6.0, line width = 0.8] (-3.25,24.75) rectangle node {\large X,H} (-0.75,21.5);
        \draw [fill={rgb,255:red,159; green,234; blue,219}, rounded corners=6.0, line width = 0.8] (6.75,24.75) rectangle node {\large X,H} (9.25,21.5); 

        \draw [fill={rgb,255:red,159; green,234; blue,219}, rounded corners=6.0, line width = 0.8] (8.25,21) rectangle node {\small Internal memory} (11.25,20.5); 
        \draw [fill={rgb,255:red,159; green,234; blue,219}, rounded corners=6.0, line width = 0.8] (-3.5,21) rectangle node {\small Internal memory} (-0.5,20.5);

        \draw [short] (-2.75,23.75) -- (-2.75,22.25);
        \draw [short] (-1.25,23.75) -- (-1.25,22.25);
        \draw [line width=1.2pt, fill = {rgb,255:red,255; green,175; blue,122}]  (-2.25,22.5) rectangle node {\normalsize $\hat{\mathrm{O}}$} (-1.75,22);
        \draw [short] (-2.75,22.25) -- (-2.25,22.25);     
        \draw [short] (-1.25,22.25) -- (-1.75,22.25);

        \draw [->, >=Circle] (-2.75,23.75) -- (-3.315,23.75); 
        \draw [->, >=Circle] (-1.25,23.75) -- (-0.75+0.065,23.75);
        \draw [->, >=Circle] (-2,22) -- (-2,21.5-0.065);
        \draw [->, >=Circle] (-2,21.5) -- (-2,21-0.065);

        \node [font=\small] at (-2,25) {Quantum Processor};

        \draw [short] (7.25,23.75) -- (7.25,22.25);
        \draw [short] (8.75,23.75) -- (8.75,22.25);
        \draw [line width=1.2pt, fill = {rgb,255:red,255; green,175; blue,122}] (7.75,22.5) rectangle node {\normalsize $\hat{\mathrm{O}}$} (8.25,22);
        \draw [short] (8.75,22.25) -- (8.25,22.25);

        \draw [->, >=Circle] (7.25,23.75) -- (6.75-0.065,23.75);
        \draw [->, >=Circle] (8.75,23.75) -- (9.25+0.065,23.75);

        \node [font=\small] at (8,25) {Quantum Processor};
        \draw [fill={rgb,255:red,159; green,234; blue,219}, rounded corners=11.3, line width = 0.8] (-5,24.85) rectangle (-4.25,23.35); 
        \draw [<->, >=Latex] (-4.625,24.5) -- (-4.625,24.1); 
        \node [font=\small] at (-4.625,23.75) {$\left|0\right\rangle$};
        \node [font=\normalsize] at (-4.625,23.1) {SPS}; 

        \draw [short, line width=.8pt] (-4.9,24.5) -- (-4.35,24.5); 
        \draw [short, line width=.8pt] (-4.9,24.1) -- (-4.35,24.1);

        \draw [<->, >=Circle] (-4.25-0.065,23.75) -- (-3.25+0.065,23.75);
        \draw [<->, >=Circle] (-0.5-0.065,20.75) -- (0.75+0.065,20.75);
        \draw [->, >=Circle] (0,23.75) -- (-0.75-0.065,23.75);
        \draw [short] (0,23.75) -- (0,24.25);
        \draw [->, >=Circle] (0,24.25) -- (0.75+0.065,24.25);
        \draw [->, >=Circle] (6.15,23.75) -- (6.75+0.065,23.75);
        \draw [short] (6.15,23.75) -- (6.15,24.25);
        \draw [->, >=Circle] (6.15,24.25) -- (5.5-0.065,24.25);
        \draw [short] (7.25,22.25) -- (7.75,22.25);

        \draw [<->, >=Circle] (9.25-0.065,23.75) -- (10+0.065,23.75);
        \draw [<->, >=Circle] (10.6,22.25+0.065) -- (10.6,21-0.065);
        \draw [<->, >=Circle] (8.25+0.065,20.75) -- (5.5-0.065,20.75);

        \draw [fill={rgb,255:red,159; green,234; blue,219}, rounded corners=10.8, line width = 0.8] (10,24.25) rectangle (11.2,22.25);
        \draw [->, >=Circle] (10.6,23.75) -- (10-0.065,23.75);
        \draw [short] (10.6,23.75) -- (10.6,23.25);
        \draw [->, >=Circle] (10.6,22.75) -- (10.6,22.25-0.065);

        \node [font=\normalsize] at (-2.375,25.75) {\textbf{ALICE}};
        \node [font=\normalsize] at (8.625,25.75) {\textbf{BOB}};

        \node [font=\footnotesize] at (-3.55,23.55) {qin};
        \node [font=\footnotesize] at (-0.32,23.55) {qout};
        \node [font=\footnotesize] at (6.45,23.55) {qin};
        \node [font=\footnotesize] at (9.25,23.55) {qout};
        \node [font=\footnotesize] at (10.3,24) {cin};
        \node [font=\footnotesize] at (10.95,22.1) {cout};
        \node [font=\small] at (3.12,21.1) {Classical Channel};
        \draw [fill={rgb,255:red,255; green,153; blue,153}, rounded corners=6.0, line width = 0] (2.75,26) rectangle node {\footnotesize \textbf{DCR}} (4.25,25.25); 
        \draw [fill={rgb,255:red,255; green,153; blue,153}, rounded corners=6.0,line width = 0] (1,23) rectangle node {\footnotesize \textbf{FIBER LOSS}} (3.7,22.5);

        \node [font=\small] at (10.6,24.8) {Quantum};
        \node [font=\small] at (10.6,24.45) {Detector};
        \node[meter, scale = 0.5, fill={rgb,255:red,255; green,175; blue,122}] (meter) at (10.6,23) {};
    \end{circuitikz}
}
  \caption{Schematic representation of the simulated setup for implementing a prepare-and-measure discrete-variable QKD protocol \cite{Martin_Megino_Non-ideal-QKDNs_2025}. This figure illustrates all the components involved in the quantum communication process, most of which are implemented using the \texttt{NetSquid} Python library \cite{coopmans2021netsquid}. Both nodes are equipped with a quantum processor for qubit state manipulation. The sender node includes a qubit source functioning as a single-photon source, while the receiver node features a quantum measurement unit (quantum detector) acting as a single-photon detector.}%
  \label{fig:NetSquid}%
\end{figure*}

The simulation setup introduced in this work can be found in the following GitHub repository \cite{Martin_Megino_Non-ideal-QKDNs_2025}. It represents the first main contribution of this work, and it was designed to simulate the quantum communication process of a generic prepare-and-measure discrete-variable QKD (DV-QKD) protocol. In such protocols, and under ideal conditions, Alice uses a single-photon source (SPS) that emits photons with a repetition time $s$. She encodes classical bits into photonic quantum states using a discrete variable (typically the  polarization) and sends them through a quantum channel. Bob, synchronized with Alice, activates his single-photon detector (SPD) during time windows of duration $\Delta t$ to measure the incoming photons. The measurement outcomes are stored as classical bits in Bob’s memory, forming the shared secret key.
The security of DV-QKD protocols is guaranteed by the Heisenberg uncertainty principle, as any attempt to measure a quantum state unavoidably disturbs it, enabling the legitimate users to detect the presence of the eavesdropper.
However, we also intend to simulate this kind of protocols in real-world scenarios. That is why our simulation setup incorporates all the non-idealities mentioned earlier in this section, distinguishing between the emission, transmission, and detection stages of the protocol. While experimental QKD implementations involve numerous side effects (see \cite{ETSI_GS_QKD_003_V2.1.1} for more details), we limit our scope to a selected subset of factors. This allows us to strike a balance between realism and simplicity, capturing the most relevant imperfections while keeping the model manageable.

To implement this approach, we used the \texttt{NetSquid} Python library. \texttt{NetSquid} is a discrete-event simulation platform designed to model all aspects of quantum networks and modular quantum computing systems, including every layer of the typical QKDN architecture \cite{coopmans2021netsquid}. Our focus is on the two lowest layers: the key manager and the physical layer. A schematic representation of a basic QKD link at the physical layer is shown in Fig.~\ref{fig:NetSquid}.
 \texttt{NetSquid} is a widely-used, powerful and versatile tool for simulating quantum networks. Its qubit-based framework facilitates the simulation of complete network architectures, including interactions between quantum and classical components. This abstraction makes it especially useful for modeling protocols and network behavior at a high level. However, this same abstraction also presents a limitation: qubits, which are idealized representations of two-level systems. By restricting quantum communication to them, \texttt{NetSquid} cannot directly simulate more physically realistic states, such as coherent states of light. Expanding support for such states would improve the realism of DV-QKD simulations and extend the range of protocols that can be studied. For example, continuous-variable (CV) QKD protocols \cite{grosshans2002continuous}, which encode information in the quadratures of the electromagnetic field. While this limitation is beyond the scope of the present work, it highlights an important direction for future research.

Let us now explain all the non-idealities we consider in our simulation framework, along with other relevant technical details.
First, we summarize the performance of the emission and detection stages defining the emission efficiency $\eta_e$  and the detection efficiency $\eta_d$ as follows:
\begin{equation}\label{emission_eff}
    \eta_e = P(\text{photon transmitted} \mid \text{photon generated}),
\end{equation}
\begin{equation}\label{detection_eff}
    \eta_d = P(\text{photon measured} \mid \text{photon received}).
\end{equation}
The emission efficiency of a SPS based on faint laser pulses can be characterized as $\eta _e = 1-e^{-\mu}$, where $\mu$ is the mean photon number per pulse. $\mu = 0.1$ is a typical value \cite{lounis2005single}.
We also consider dark counts in the detection stage. The probability of a dark count occurring during the time window $\Delta t$ while the detector is active is given by
\begin{equation}\label{P_DCR}
    P_{DCR}=1-e^{-R_{DCR}\cdot\Delta t},
\end{equation}
where $R_{DCR}$ is the dark count rate, measured in $\mathrm{s}^{-1}$.

To model the transmission stage, we first define qubit loss in the quantum channel based on the Beer-Lambert law, expressing the probability of photon loss as a function of the channel length $d$:
\begin{equation}\label{P_loss}
    P_{loss}=1-(1 - P_0) \cdot 10^{-\frac{R}{10}d}.
\end{equation}
Here, $P_0 := 1-\eta_{e}\eta_{d}$ represents the probability of photon loss occurring during the emission or detection stages, while $R$ denotes the photon loss rate, measured in $\mathrm{dB/km}$.
On the other hand, as photons propagate through the channel, they interact with the medium and become susceptible to environmental noise. In this work, we adopt the depolarizing noise model \cite{wilde2013quantum} (see also \cite[p.~178]{nielsen2010quantum}), where a photon initially encoded in the pure state $\left|s\right\rangle$ is received at the end of the channel as
\begin{equation}\label{P_D}
    \rho=(1-P_{depolar})\left|s\right\rangle \left\langle s\right|+P_{depolar}\frac{\mathbb{I}}{2},
\end{equation}
where $P_{depolar}$ represents the probability that the qubit's quantum information is completely lost and replaced by a maximally mixed state. This quantity is given by
\begin{equation}
    P_{depolar}=1-e^{-\frac{R_{depolar}}{v_{f}}d},
\end{equation}
where $R_{depolar}$ is the depolarizing rate (measured in $\mathrm{s}^{-1}$), and $v_{f}$ is the transmission speed in the quantum channel.

\begin{figure}[t!]%
\resizebox{0.48\textwidth}{!}{%
\begin{circuitikz} 
\tikzstyle{every node}=[font=\large] 
\draw [line width=1.5pt]  (-26.25,29.75) rectangle  node {\LARGE \textbf{ALICE}} (-23.75,28.5); 
\draw [line width=1.5pt] (-15,29.75) rectangle  node {\LARGE \textbf{BOB}} (-12.5,28.5); 

\draw [dashed, line width=1pt] (-26.25,27.25) -- (-12.5,27.25); 
\node [font=\LARGE] at (-11.65,27.25) {$t=0$}; 
\draw [->, >=Stealth, line width=1pt] (-27,27.25) -- (-27,4.75); 
\node [font=\LARGE, rotate around={90:(0,0)}] at (-27.5,7.25) {\LARGE Time $t$}; 
\draw [dashed, line width=1pt] (-25,28.5) -- (-25,4.75); 
\draw [dashed, line width=1pt] (-13.75,28.5) -- (-13.75,4.75); 

\draw[-Stealth, line width = 2pt, color = {rgb,255:red,236; green,219; blue,34}] (-13.75,19.75) to[bend right=70] (-13.75,13.5);
\draw[line width = 1pt, color = {rgb,255:red,236; green,219; blue,34}] (-13.75,18.5) to[bend right=70] (-13.75,14.75);
\draw[line width = 1pt, color = {rgb,255:red,236; green,219; blue,34}] (-13.75,21) to[bend right=70] (-13.75,12.25);
\fill [color={rgb,255:red,236; green,219; blue,34}, opacity=0.2]  (-13.75,21) -- (-13.75,18.5) to[bend right=70] (-13.75,14.75) -- (-13.75,12.25) to[bend left=70] (-13.75,21); 
\fill [pattern=north east lines,pattern color={rgb,255:red,236; green,219; blue,34}]  (-13.75,21) -- (-13.75,18.5) to[bend right=70] (-13.75,14.75) -- (-13.75,12.25) to[bend left=70] (-13.75,21); 

\draw [color = {rgb,255:red,143; green,143; blue,143}, line width=0.5pt] (-25,23.5) -- (-13.75,21); 
\draw [color = {rgb,255:red,143; green,143; blue,143}, line width=0.5pt] (-25,23.5) -- (-13.75,18.5); 
\fill [fill opacity = 0.75, fill={rgb,255:red,255; green,255; blue,255}] (-25,23.5) -- (-13.75,21) -- (-13.75,18.5) -- (-25,23.5) -- cycle;
\fill [pattern=north west lines,pattern color={rgb,255:red,143; green,143; blue,143}] (-25,23.5) -- (-13.75,21) -- (-13.75,18.5) -- (-25,23.5) -- cycle;

\draw [ color={rgb,255:red,145; green,14; blue,216}, line width=3pt, Bracket - Bracket] (-25,27.25) -- (-25,23.5); 
\draw [->, >=Stealth, line width=1.5pt] (-25,23.5) -- (-13.75,19.75); 
\draw [<->, >=Stealth, line width=1.5pt] (-23.75,27.25) -- (-23.75,25.25); 
\draw [<->, >=Stealth, line width=1.5pt] (-23.75,25.25) -- (-23.75,23.5); 
\node [font=\LARGE] at (-22,26.25) {Bit encoding}; 
\node [font=\LARGE] at (-21.75,24.25) {Basis encoding}; 

\coordinate (Mid) at ($(-25,23.5)!0.5!(-13.75,19.75)$);
\coordinate (MidDown) at ($(Mid) + (0,-3)$);

\coordinate (Vec) at ($(-13.75,19.75) - (-25,23.5)$);         
\coordinate (HalfVec) at ($0.15*(Vec)$);                       
\coordinate (Start) at ($(MidDown) - (HalfVec)$);
\coordinate (End) at ($(MidDown) + (HalfVec)$);
\draw [->, >=Stealth, line width=1.5pt] (Start) -- (End);
\node[above] at (MidDown) {\LARGE $v_f$};
\coordinate (W) at ($(MidDown) - 0.5*(HalfVec)$);
\coordinate (E) at ($(MidDown) + 0.5*(HalfVec)$);
\coordinate (N) at ($(W) + (1.6875, 0)$);            
\coordinate (S) at ($(W) + (0, -0.5625)$);            

\draw[dashed, line width=0.7pt] (W) -- (S);
\draw[dashed, line width=0.7pt] (S) -- (E);
  
\node [font=\LARGE, rotate around={90:(0,0)}] at (-25.35,25.25) {\LARGE $2\cdot \mathrm{GD}_A$}; 
\draw [ color={rgb,255:red,39; green,169; blue,30}, line width=3pt, Bracket - Bracket] (-13.75,17.25) -- (-13.75,22.25); 
\draw [dashed, line width=1pt] (-13.75,21) -- (-11.25,21); 
\draw [dashed, line width=1pt] (-13.75,18.5) -- (-11.25,18.5); 
\draw [<->, >=Stealth, line width=1.5pt] (-11.25,21) -- (-11.25,18.5); 
\node [font=\LARGE, rotate around={90:(0,0)}] at (-11.6,19.75) {$2\delta\tau$}; 
\draw [<->, >=Stealth, line width=1.5pt] (-10,22.25) -- (-10,17.25); 
\node [font=\LARGE, rotate around={90:(0,0)}] at (-10.35,19.75) {$2C\delta\tau$}; 
\draw [dashed, line width=1pt] (-13.75,22.25) -- (-10,22.25); 
\draw [dashed, line width=1pt] (-13.75,17.25) -- (-10,17.25); 
\draw [<->, >=Stealth, line width=1.5pt] (-12.5,27.25) -- (-12.5,22.25); 
\node [font=\Large, rotate around={90:(0,0)}] at (-12.9,24.75) {$2\cdot \mathrm{GD}_A+\tau-C\delta\tau$}; 
\draw [dashed, line width=1pt] (-14,19.75) -- (-12.5,19.75); 
\draw [<->, >=Stealth, line width=1.5pt] (-12.5,19.75) -- (-12.5,16.25); 
\node [font=\LARGE, rotate around={90:(0,0)}] at (-12.9,17.9) {$\mathrm{GD}_B$}; 
\draw [<->, >=Stealth, line width=1.5pt] (-12.5,16.25) -- (-12.5,13.5);

\draw [<->, >=Stealth, line width=1.5pt] (-12.5,13.5) -- (-12.5,11.5); 
\node [font=\LARGE, rotate around={90:(0,0)}] at (-12.85,12.65) {$\mathrm{DT}$}; 
\node [font=\LARGE, rotate around={90:(0,0)}] at (-12.85,14.85) {$\mathrm{DD}$}; 
\draw [->, >=Stealth, line width=1pt] (-25,27.75) -- (-13.75,27.75); 
\node [font=\LARGE] at (-19.5,28.15) {\LARGE Distance $d$}; 

\draw [ color={rgb,255:red,39; green,169; blue,30}, line width=3pt, Bracket - Bracket] (-13.75,11) -- (-13.75,16); 
\draw [dashed, line width=1pt] (-13.75,16.25) -- (-12.5,16.25); 
\draw [dashed, line width=1pt] (-13.75,13.5) -- (-10,13.5); 
\draw [dashed, line width=1pt] (-13.75,11.5) -- (-12.5,11.5); 
\draw [ color={rgb,255:red,39; green,169; blue,30}, line width=3pt, - Bracket] (-13.75,4.75) -- (-13.75,9); 
\draw [ color={rgb,255:red,39; green,169; blue,30} , fill={rgb,255:red,39; green,169; blue,30}] (-13.75,19.75) circle (0.15cm); 
\draw [ color={rgb,255:red,39; green,169; blue,30} , fill={rgb,255:red,39; green,169; blue,30}] (-13.75,13.5) circle (0.15cm); 

\draw [color = {rgb,255:red,143; green,143; blue,143}, line width=0.5pt] (-25,10.25) -- (-13.75,7.75); 
\draw [color = {rgb,255:red,143; green,143; blue,143}, line width=0.5pt] (-25,10.25) -- (-13.75,5.25); 
\fill [pattern=north west lines, pattern color={rgb,255:red,143; green,143; blue,143}] (-25,10.25) -- (-13.75,7.75) -- (-13.75,5.25) -- (-25,10.25) -- cycle;
\draw [->, >=Stealth, line width=1.5pt] (-25,10.25) -- (-13.75,6.5); 

\draw [ color={rgb,255:red,145; green,14; blue,216}, line width=3pt, Bracket - Bracket] (-25,14) -- (-25,10.25); 
\node [font=\LARGE, rotate around={90:(0,0)}] at (-25.4,12) {$2\cdot \mathrm{GD}_A$}; 
\draw [ color={rgb,255:red,237; green,12; blue,12}, line width=2pt, <->, >=Stealth] (-26,27.25) -- (-26,14); 
\node [font=\LARGE, rotate around={90:(0,0)}] at (-26.25,20) {$s$}; 
\draw [dashed, line width=1pt] (-26,14) -- (-25,14); 
\node [font=\LARGE, rotate around={90:(0,0)}] at (-10.35,10) {$2C\delta\tau + \mathrm{DT}$};
\draw [<->, >=Stealth,  line width=1.5pt] (-10,13.5) -- (-10,6.5); 
\draw [dashed, line width=1pt] (-25,23.5) -- (-23.75,23.5); 
\draw [dashed, line width=1pt] (-25,25.25) -- (-23.75,25.25); 
\draw [dashed, line width=1pt] (-25,10.25) -- (-11.25,10.25); 
\draw [dashed, line width=1pt] (-13.75,6.5) -- (-10,6.5); 
\draw [<->, >=Stealth, line width=1.5pt] (-11.2,10.25) -- (-11.25,6.5); 
\node [font=\LARGE, rotate around={90:(0,0)}] at (-11.6,8.5) {$\tau$}; 
\draw [ color={rgb,255:red,39; green,169; blue,30} , fill={rgb,255:red,39; green,169; blue,30},] (-13.75,6.5) circle (0.15cm);

\draw[fill={rgb,255:red,255; green,255; blue,255}, fill opacity = 0.85, text opacity = 1]  (-26,6.25) rectangle  node {\LARGE $s_{lim}=\max\left[2GD_{A},\mathrm{DD+GD_{B}+\mathrm{DT}+2C\delta\tau}\right]$} (-14.25,5);
\end{circuitikz} 
}%
\caption{Time-space diagram showing the emission, reception, and measurement of a single photon, used to determine the repetition time $s$ for Alice's SPS in the BB84 protocol. Violet segments represent the time needed to manipulate the emitted photon's state, gray dashed areas represent the channel's timing jitter $\delta\tau$, green segments indicate the time windows $\Delta t=C\delta\tau$ (where $C=3$) of detector activation, and the yellow area shows the time from photon arrival to detection. The diagram provides an appropriate value for $s$, which is not the minimum for correct synchronization but is defined based on the actual lower bound. The relative sizes of the intervals are not to scale.}
\label{fig:Synch}
\end{figure}

These theoretical models characterize the non-ideal behavior of the complete quantum transmission process. However, in order to successfully execute QKD protocols we had to address other technical issues common in experimental setups, such as timing jitter. This includes random variations in the arrival time of photons during transmission, which we model as a Gaussian random variable with a mean value of $\tau :=d/ v_f$ and standard deviation $\delta \tau :=\mathrm{std}\cdot\tau$, where $\mathrm{std}$ is a fixed parameter. We also simulated the timing jitter of the single-photon detector (SPD), which represents the delay between photon arrival and measurement \cite{lee2023quantum}. This is modeled by setting a value for the detection delay parameter ($\mathrm{DD}$) in Bob's detection unit. Additionally, we accounted for the detector's dead time ($\mathrm{DT}$), the period immediately following a photon detection during which the detector is unable to register subsequent events.
These timing constraints raise a second critical concern: synchronization between Alice and Bob. As previously mentioned, Bob activates his detector for a specific period, $\Delta t$. Proper synchronization ensures that Bob correlates this time window with the expected photon arrival time and adjusts the window width based on the channel's timing jitter, applying a coverage factor $C=3$ such that $\Delta t = C \cdot \delta \tau$. Moreover, synchronization of photon emission and detection sets a lower bound, $s_{lim}$, for the emission repetition time, $s$. Alice selects her value of $s$ based on this lower bound:
\begin{equation}
\resizebox{\columnwidth}{!}{$
\begin{aligned}[t]
s &= \max\!\left[3\mathrm{GD}_{A},\, \mathrm{DD} + \mathrm{GD}_{B} + \mathrm{DT} + 3C\delta\tau\right] \\[6pt]
  &\geq s_{\mathrm{lim}} = \max\!\left[2\mathrm{GD}_{A},\, \mathrm{DD} + \mathrm{GD}_{B} + \mathrm{DT} + 2C\delta\tau\right]
\end{aligned}
$}
\end{equation}
where $\mathrm{GD}_A$ and $\mathrm{GD}_B$ are the gate duration of Alice's and Bob's quantum processors, respectively. This formula should take into account the physical limitations of the device, as real SPSs used for QKD typically operate at repetition rates on the order of 1 GHz \cite{pirandola2020advances}. However, for the sake of simplicity, these considerations have been omitted in this work.
In order to properly synchronize with each other, Alice and Bob need information about some characteristic parameters of the channel and their devices, as shown above. See Fig.~\ref{fig:Synch} for an explanation of the synchronization process and the calculation of $s_{lim}$.

\renewcommand*\arraystretch{1.2}
\begin{table}[H]
\begin{centering}
\resizebox{\columnwidth}{!}{%
\begin{tabular}{|>{\centering}p{1.9cm}|>{\centering}p{1.9cm}||>{\centering}p{1.9cm}|>{\centering}p{1.9cm}|}
\hline 
Parameter & Value & Parameter & Value\tabularnewline
\hline 
$\eta_{e}$ & 0.2 & $R_{DCR}$ & 25 $\mathrm{s}^{-1}$\tabularnewline
\hline 
$\eta_{d}$ & 0.6 & $\mathrm{DD}$ & 0.5 ns \tabularnewline
\hline 
$R$ & 0.2 dB/km & $\mathrm{DT}$ & 100 ns \tabularnewline
\hline 
$v_{f}$ & $2/3\cdot c$ & $\mathrm{GD}_{A}$ & 1 ns\tabularnewline
\hline 
$\mathrm{std}$ & 0.02 & $\mathrm{GD}_{B}$ & 1 ns\tabularnewline
\hline 
$R_{depolar}$ & 100 $\mathrm{s}^{-1}$ & ~ & ~ \tabularnewline
\hline
\end{tabular}
}
\par\end{centering}
\caption{Numerical values used for protocol simulation.}
\label{T1}
\end{table}

Having covered all relevant technical details of the simulation setup, in Tab.~\ref{T1} we provide numerical values for some of the parameters defined above. These values, which will remain constant in our simulations, are chosen based on typical experimental implementations \cite{lounis2005single, eisaman2011invited, pirandola2020advances, chen2021integrated}, though they may not fully represent real devices.

\section{Variable-Length BB84 Protocol}\label{3}
The origins of QKD trace back to the 1970s, when researchers began exploring quantum phenomena for secure communication. The landmark development in this field came in 1984 with the introduction of the BB84 protocol by Charles Bennett and Gilles Brassard \cite{BEN84}. This protocol not only demonstrated the theoretical feasibility of QKD but also laid the foundation for practical implementations that continue to evolve today. BB84 exploits quantum superposition and the uncertainty principle to establish a shared secret key while enabling the detection of eavesdropping through the measurement postulate.
In their original work, the authors proposed using photon polarization as qubits. A photon’s polarization describes the orientation of its electromagnetic wave oscillations and can be represented using two orthogonal basis states. We will denote them as $\left|0\right\rangle$ (the electric field oscillates horizontally) and $\left|1\right\rangle$ (the electric field oscillates vertically), to remark the analogy with the classical bit states ``0" and ``1". Due to the quantum nature of polarization, photons can exist in superposition states. Specifically, the diagonal polarization states, defined as $\left|\pm\right\rangle =\frac{1}{\sqrt{2}}\left(\left|0\right\rangle \pm\left|1\right\rangle \right)$, further expand the encoding possibilities within the BB84 protocol.

In the BB84 protocol, Alice and Bob are connected by a two-way classical channel and a one-way quantum channel. After authenticating each other and synchronizing their devices, they execute the protocol, which consists of two main phases.
The first, denoted here as the quantum phase, involves Alice generating a random key and transmitting it to Bob by encoding the classical information into quantum states. The second, the post-processing phase, includes applying error correction techniques to compensate for transmission errors and ensure the secrecy of the final key. 

We propose a variable-length version of BB84. In this class of protocols, as defined in~\cite{tupkary2024security}, the output key size is not fixed beforehand but is dynamically determined during execution.
In our approach, the quantum phase is executed once, with the number of photons estimated in advance to meet the desired output length in a single round. We opted for this strategy instead of block-based alternatives, which repeatedly send fixed-size photon blocks until the raw key reaches a sufficient size. This introduces extra variability in the final output length and an uncontrolled risk of falling short of the target, particularly at short distances. Additional rounds incur time overhead and reduce temporal predictability. Moreover, when each block yields few raw bits, QBER estimation can become unreliable, undermining security guarantees. In contrast, a single-round approach improves efficiency and reduces execution time uncertainty, avoiding the need to rerun the quantum phase. We also implemented and tested a block-based approach for comparison. The results, not shown here for brevity, confirm that the single-round method provides better alignment with output length targets and more reliable noise estimation across a wide range of distances.
Our version of BB84 is designed to support predefined output length requirements while ensuring reliable parameter estimation and successful execution of post-processing steps. To achieve this, we introduce an additional step in the quantum phase: a controlled bit-flip mechanism applied locally by Bob to his version of the raw key. This step, inspired by earlier proposals \cite{renner2008security}, can be understood as the deliberate introduction of extra noise. It is simple to implement, does not compromise the protocol's security, and improves the predictability of the information reconciliation process. As we will show later, it also enables a more efficient execution by keeping the final key length above and closely aligned with the user’s specified target.
In addition, prior knowledge of certain channel parameters is still required for operations such as synchronization and post-processing. 
This version of the BB84 protocol operates as follows:
\subsection*{Quantum phase}
\begin{enumerate}[itemsep=1pt, labelindent=0pt, itemindent=0pt, labelwidth=1em, labelsep=0.5em, leftmargin=1.5em]
    \item Alice generates a random $N$-bit string, $k_A$.\footnote{Random bit choices are equiprobable: $P(0) = P(1) = 1/2$.}
    \item Alice and Bob each generate random bit strings, $b_A$ and $b_B$, to select their encoding bases. 
\end{enumerate}

\begin{enumerate}[start = 3, itemsep=1pt, labelindent=0pt, itemindent=0pt, labelwidth=1em, labelsep=0.5em, leftmargin=3.5em]
    \item Alice encodes the key into photonic polarization states. Each photon, initially in state $\left|0\right\rangle$, is manipulated with quantum gates based on the corresponding bits $k_{A,i}$ and $b_{A,i}$, producing the state $\left|\Psi_i\right\rangle = H^{b_{A,i}} X^{k_{A,i}} \left|0\right\rangle$.\footnote{The $i$-th component of an array $V_X$ is denoted as $V_{X,i}$.} $H$ is the Hadamard gate and $X$ is the first Pauli gate. $b_{A,i}=0$ corresponds to the horizontal basis in the 2D Hilbert space, and $b_{A,i}=1$ to the diagonal basis.

    \item Alice sends the state $\left|\Psi_i\right\rangle$ to Bob.
    \item Bob measures the photon polarization according to his basis choice $b_{B,i}$. Therefore, he measures the state $H^{b_{B,i}}\left|\Psi_i\right\rangle = H^{b_{A,i}\oplus b_{B,i}} X^{k_{A,i}} \left|0\right\rangle$. He records the result in $k_{B,i}$, attempting to reconstruct Alice's key. If he does not measure any signal, he accounts for that writing ``X'' in $k_{B,i}$.  
\end{enumerate}
\vspace{-9.6cm} 
\hspace*{1.6em} 
\begin{tikzpicture}[remember picture,overlay]
  \draw[decorate,decoration={brace,mirror,amplitude=6pt}, thick]
    (0,0.1) -- ++(0,-8.8)
    node[midway,xshift=-0.5cm,rotate=90] {Repeated for $i=1,...,N$};
\end{tikzpicture}

\vspace{8.2cm} 
\hspace*{-1.6em}

\begin{enumerate}[start = 6, itemsep=1pt, labelindent=0pt, itemindent=0pt, labelwidth=1em, labelsep=0.5em, leftmargin=1.5em]
    \item \textbf{Key sifting}. Alice sends $b_{A}$ to Bob via the classical channel. Bob identifies and reports to Alice the lost signals ($k_{B,i}=$``X") and the mismatched bases ($b_{A,i} \neq b_{B,i}$). Both parties remove these bits from their respective keys, resulting in two sifted keys of size $n$.
    \item \textbf{Controlled randomization phase}. Right before parameter estimation, Bob adds a predetermined amount of artificial noise to his raw key, flipping each one of its bits with probability $P_{extra}$. Then, the effective bit flip probability between Alice and Bob's raw keys is 
    \begin{equation}\label{P_flip_hat}
        \hat{P}_{flip} = P_{flip} + P_{extra} - 2\cdot P_{flip} \cdot P_{extra},
    \end{equation}
    where $P_{flip}$ is the bit flip probability corresponding to the noisy quantum channel. It is important to mention that we only consider this phase if the user introduces output length requirements. If that is not the case, $P_{extra} = 0$.

    \item \textbf{Parameter estimation}. Alice randomly selects a fraction $g(n) \leq 1/2$ of the remaining bits and sends their positions to Bob. Bob returns the corresponding bits from his key, allowing Alice to estimate the observed error rate $\hat{Q}$. The objective is to estimate the quantum bit error rate (QBER), which quantifies the errors introduced by the quantum channel. It is defined as the fraction of mismatched bits between Alice and Bob after sifting and before any controlled randomization. The true QBER, denoted $Q$, can be inferred from the observed rate $\hat{Q}$ via
    \begin{equation}
        Q = \frac{\hat{Q} - P_{\mathrm{extra}}}{1 - 2P_{\mathrm{extra}}}.
    \end{equation}
    If $Q \geq Q_t := 0.091$, the protocol is aborted. Otherwise, Alice and Bob proceed with two raw keys of length $l = (1 - g(n)) \cdot n$ and the estimated error rate $\hat{Q}$. This step helps detect potential eavesdropping.

   Note that the threshold used in this protocol for eavesdropping detection is slightly more restrictive than the commonly accepted limit for BB84, $Q \geq 0.11$ \cite{shor2000simple}. However, this assumes perfect information reconciliation, with an efficiency factor of $f = 1$. In our case, the chosen threshold $Q_t = 0.091$ corresponds to the solution of the equation $1 - (1 + f_{\max}) h(Q) = 0$, situation in which $k$ goes to 0. 
    Note also that while the inclusion of $P_{\mathrm{extra}}$ increases the uncertainty in estimating the QBER, its value is chosen in advance based on the link parameters and is set to zero when operating near the security threshold $Q_t$, as will be shown later.

\end{enumerate}


\subsection*{Post-processing}
\begin{enumerate}[itemsep=1pt, labelindent=0pt, itemindent=0pt, labelwidth=1em, labelsep=0.5em, leftmargin=1.5em]
\setcounter{enumi}{8} 
\item \textbf{Information reconciliation}. Alice and Bob use an implementation \cite{cascade} of the original Cascade protocol \cite{brassard1993secret} to correct errors in the $l$ raw key bits based on the estimated error rate $\hat{Q}$ between both raw keys. The successful execution of this protocol reveals a certain amount of key information, denoted $n_{\mathrm{exp}}$, during the reconciliation process. Although this value varies with each protocol run, it can be estimated as
\begin{minipage}{0.94\linewidth}  
\begin{equation}
    n_{exp}=f\cdot l\cdot h(\hat{Q}),
\end{equation}
\end{minipage}
where $h(x)$ is the binary entropy, and $f \gtrapprox 1$ is a parameter representing Cascade's efficiency relative to the ideal performance \cite{martinez2014demystifying}. While $n_{exp}$ is typically determined during the protocol, security standards often require using prior knowledge of the link status to calculate an upper bound based on the input bits $l$. This bound must be determined before executing the Cascade protocol \cite{tupkary2024security}. From our study of Cascade's performance in \cite{cascade}, we estimated a maximum efficiency of $f_{max} = 1.27$, which can be used to estimate an upper bound for $n_{exp}$ as 
\begin{minipage}{0.94\linewidth}  
\begin{equation}
    n_{exp}^{UB} = f_{max} \cdot l \cdot h(\hat{P}_{flip}),
\end{equation}
\end{minipage}
where $\hat{P}_{flip}$ is used instead of $\hat{Q}$.

\item \textbf{Privacy amplification}. Alice and Bob estimate $k$, a value that quantifies the remaining secrecy of their corrected keys and is related to their min-entropy. They also agree on a shared random seed. Using these, they run an implementation \cite{foreman2025cryptomite, mauerer2012modular} of the Trevisan extractor \cite{trevisan2001extractors} to transform the weakly random $l$-bit string with high min-entropy into a nearly uniform random output. The result is two identical keys, $K_A = K_B =: K_F$, of size $m < l$, where $m$ is given implicitly as \cite{foreman2025cryptomite, mauerer2012modular}
\begin{equation}\label{m}
    m=\left\lfloor k-6-4\log_{2}\left(\frac{m}{\varepsilon_{max}}\right)\right\rfloor.
\end{equation}

$\varepsilon_{max}=0.01$ is a parameter that serves as security standard.
The extractor ensures that even if the eavesdropper knows part of the corrected key after the Cascade execution, the final key remains unpredictable and secure. Technically speaking, $k$ is a lower bound of the conditional smooth min-entropy of the input bits after information reconciliation \cite{renner2008security}, $H_{min}^{\varepsilon}\left(X^{l}|CE^{l}\right)$, which accounts for the randomness of the corrected keys from an eavesdropper's point of view. This lower bound can be obtained from \cite[pp.~25,~98]{renner2008security}, \cite{scarani2008quantum} as

\vspace{-15pt}

\begin{minipage}{0.94\linewidth}    
    \begin{multline}\label{minentropy}
    H_{min}^{\varepsilon}\left(X^{l}|CE^{l}\right)\gtrapprox l\cdot H(X|E)-n_{exp}^{UB}\\
    \simeq l\cdot\left(H(X|E)-f_{max}\cdot H(X|Y)\right)\geq \\ \geq l\cdot\left(1-(1+f_{max})h(\hat{P}_{flip})\right)=:k,
    \end{multline}
\end{minipage}
where the conditional entropy involved verifies $H(X|Y)\geq1-h(\hat{P}_{flip})$ \cite{scarani2008quantum, renner2008security}. 
It is worth mentioning that Eq.~(\ref{minentropy}) is derived in the asymptotic regime, assuming that $l \longrightarrow \infty$. While we will be working with large but finite keys, this assumption is not completely realistic, and a proper security analysis that considers finite-key effects is necessary. Several approaches exist to address this issue and provide tight bounds for a secure output key length \cite{tupkary2024security, george2021numerical, tomamichel2012tight}, but such an analysis is outside the scope of this work and will be left for future studies. 
\end{enumerate}

The variability of the final key length $m$ depends solely on the randomness of photon losses and key sifting. After that and if the protocol is not aborted, the post-processing stage provides a deterministic relation between the input key length $l$ and the output length $m$, obtained joining Eqs.~(\ref{m}) and (\ref{minentropy}):
\begin{equation}\label{ml}
\resizebox{\columnwidth}{!}{$
    m=\left\lfloor l\cdot\left(1-(1+f_{max})h(\hat{P}_{flip})\right)-6-4\log_{2}\left(\frac{m}{\varepsilon_{max}}\right)\right\rfloor.
$}
\end{equation}

To execute the protocol, only three parameters need to be provided to the initialization function, which will be the focus of this work: $N$, the number of photon pulses sent by Alice; $g(\cdot)$, the fraction of raw key bits used for parameter estimation; and $P_{\mathrm{extra}}$, the artificial randomization applied by Bob to his raw key to deliberately increase the QBER. The role and purpose of $P_{\mathrm{extra}}$ will be discussed in more detail later.

\section{Analysis of Output Key Length Guarantees}\label{4}
Using the formulas from Sect.~\ref{2}, we calculate two fundamental quantities for this section. The first is the probability $p$ that Bob measures a signal within a specific time window and has a basis match for that key position. Using Eqs.~(\ref{P_DCR}) and (\ref{P_loss}), we obtain
\begin{equation}\label{p}
    p = \frac{1}{2} \left( 1 - P(X) \right) = \frac{1}{2} \left( 1 - P_{loss} \left( 1 - P_{DCR} \right) \right),
\end{equation}
where $P(X)=P_{loss} \left( 1 - P_{DCR} \right)$ is the probability of Bob not measuring any signal at all within a time window.
This probability $p$ defines the expected number of bits that remain after the sifting phase: $n=Np$.
We can also calculate the expected QBER as the probability $P_{flip}$ of having classical bit flip between $k_{A,i}$ and $k_{B,i}$ right after the quantum transmission phase and before controlled randomization, assuming that this $i-$th bit survived the sifting phase (a signal was measured and bases coincided for the $i$-th time window). Considering Eqs.~(\ref{P_DCR}), (\ref{P_loss}) and (\ref{P_D}), this probability can be calculated as
 \begin{equation}\label{Pflip}
 \resizebox{\columnwidth}{!}{$
  \text{\small
  $
  P_{flip} = \frac{\frac{P_{DCR}}{4}\left(P_{loss}+1\right)+\frac{P_{depolar}}{2}\left(1-\frac{1}{2}P_{DCR}\right)\left(1-P_{loss}\right)}{1-P_{loss}\left(1-P_{DCR}\right)}
  $}.
  $}
\end{equation}
See Appendix~\ref{appA} for additional details on the derivation of Eq.~(\ref{Pflip}). 

Having established the complete theoretical framework of our BB84 protocol,  let us address a reversed analysis of the problem, which is the cornerstone of this work. To illustrate the need for flexible interaction between KMs and QKD modules in a QKDN, able to adapt to several scenarios, consider the following example. Suppose Alice and Bob are connected by a QKD link with certain characteristics (length $d$) and need to establish a shared secret key of size $m_F$ for secure communication over the classical channel. Typically, KMs would provide keys stored from previous QKD executions. However, if their shared key buffer is empty, they must rely on the QKD modules to generate at least $m_F$ bits to meet the application layer's requirements in the shortest time possible. In such situation, the question is: what is the minimum number of photons $N$ that Alice must send in the initial stage of the protocol to ensure that a secret key of size $m\geq m_F$ is generated with high probability? In other words, we want to determine the values of initial single-photon pulses $N$ such that
 \begin{equation}\label{condition1}
     \left\langle M(N)\right\rangle -C_F \cdot \sigma_M (N)\geq m_F,
 \end{equation}
 where $M$ is the output key length interpreted as a random variable. In fact, apart from the variable $N$ and a few other exceptions, we refer to random variables using capital letters. The mean value will be denoted as $\left\langle M\right\rangle$, and the standard deviation as $\sigma_M$. The parameter $C_F$ represents the confidence level in terms of $\sigma_M$ for obtaining at least $m_F$ secret bits. We set this value to $C_F = 3$.

To address the previous question, we try to estimate $\sigma_M$ in terms $N$. First, we will try to rewrite Eq.~(\ref{condition1}) in terms of the length of the raw key, $L$. Neglecting non-linear terms in $m$, it follows from Eq.~(\ref{ml}) that
 \begin{equation}\label{deltam}
     \sigma_M \simeq\left(1-(1 + f_{max})h(\hat{P}_{flip})\right)\sigma_L.
 \end{equation}
Using Eq.~(\ref{deltam}) and Eq.~(\ref{ml}), Eq.~(\ref{condition1}) is equivalent to
\begin{equation}\label{condition2}
     \left\langle L(N)\right\rangle-C_F \cdot \sigma_L (N)\geq l_F,
 \end{equation}
where 
 \begin{equation}\label{lf}
    l_{F}=\frac{m_{F}+6+4\log_{2}\left(m_{F}/\varepsilon_{max}\right)}{1-(1+f_{max})\cdot h(\hat{P}_{flip})}
 \end{equation}
is the key length required after the quantum phase to obtain $m_F$ bits in the final key. To determine $\left\langle L\right\rangle$ and $\sigma_L$ in terms of $N$ and other characteristic parameters, we first need to define a strategy for allocating a fraction of bits, $g(n)$, to parameter estimation in the quantum phase. This is a crucial consideration, as there must be a balance between achieving an accurate estimation $\hat{Q}$ of the effective QBER, and retaining a sufficient fraction of bits for post-processing. Given that $n$ follows a binomial distribution, $n\equiv Y \sim\mathrm{Bin}(N,p)$, we define two random variables: the number of bits used for parameter estimation, $K = g(Y) \cdot Y$, and the remaining key bits, $L = (1-g(Y)) \cdot Y$. Assuming we are working with sufficiently large values, we can approximate the distribution of our error rate estimation $\hat{Q}$ using a normal distribution (see Appendix~\ref{appB} for extra details) to find
\begin{equation}\label{Q}
    \hat{Q}\simeq\mathcal{N}\left(\hat{P}_{flip},\frac{\hat{P}_{flip}\left(1-\hat{P}_{flip}\right)}{\mathbb{E}\left[g(Y)\cdot Y\right]}\right),
\end{equation}
where $\hat{P}_{flip}$ is defined in Eq.~(\ref{P_flip_hat}). To measure the accuracy in the estimation of the QBER, we propose the following condition:
 \begin{equation}\label{condition3}
     \sigma_{\hat{Q}}/{\left\langle \hat{Q}\right\rangle }\leq \Gamma(\varepsilon):=\varepsilon(1+\alpha\cdot10^{-\beta \hat{P}_{flip}}),
 \end{equation}
where $\varepsilon=0.1$, $\alpha=3$ and $\beta=20$. However, this parameter selection is not fixed and can be set as desired. An accurate estimation of the QBER is essential not only for eavesdropping detection but also for the proper execution of Cascade. This interactive error correction method compares block parities between Alice's and Bob's keys, starting with a block length inversely proportional to the expected QBER. It is therefore important to have a sufficient number of bits at this stage to ensure that $\hat{Q}$, $\hat{P}_{flip}$, and the actual error rate of the raw key are closely aligned. The condition proposed in Eq~(\ref{condition3}) allows for a less restrictive condition when $\hat{P}_{flip}$ is closer to 0, balancing an effective QBER estimation in low-error scenarios with achievable lower bounds for the number of initial photons.  

In this work, we have considered three different parameter estimation strategies, denoted as $g_i(n)$, each defined by a distinct function. The first two strategies, $i = 1$ and $i = 2$, are based on a constant fraction and a constant number of raw key bits, respectively. These approaches are simple to implement, making them useful baselines for comparison. The third strategy, proposed in \cite{scarani2008quantum}, is included because it is suggested by the authors to be optimal in terms of the trade-off between parameter estimation accuracy and key rate. By analyzing all three, we aim to capture both practical and theoretically grounded perspectives in our evaluation.
For each of these functions, we have estimated the mean value $ \left\langle L \right\rangle$ and the standard deviation $\sigma_L$ as functions of $N$. We have also adapted Eq.~(\ref{condition3}) to each specific case, since we know it is equivalent to
\begin{equation}
    \mathbb{E}\left[g(Y)\cdot Y\right]\geq A_0:=\frac{1}{\Gamma(\varepsilon)^2}\left(\frac{1}{\hat{P}_{flip}}-1\right).
\end{equation}
A summary of the results is presented in Tab.~\ref{tab3}:

\begin{table}[h]
\renewcommand*\arraystretch{2.2}
\begin{centering}
\resizebox{\columnwidth}{!}{%
{\footnotesize
\begin{tabular}{|P{1.8cm}|P{2cm}|P{2cm}|P{1.8cm}|}
\hline 
{$g_{i}(n)$} & {$\left\langle L\right\rangle$} & {${\displaystyle \frac{\sigma_L}{\sqrt{Np\left(1-p\right)}}}$} & {${\displaystyle \frac{\sigma_{\hat{Q}}}{\left\langle \hat{Q}\right\rangle }\leq \Gamma(\varepsilon)}$} \tabularnewline
\hhline{|====|} 
\parbox[c]{2cm}{\centering $g_{1}(n) = g$\\ $\in (0,1)$} & {$\left(1-g\right)Np$} & {$\left(1-g\right)$} & {${\displaystyle N\geq\frac{A_{0}}{p\cdot g}}$} \tabularnewline
\hline 
\parbox[c]{2cm}{\centering ${\displaystyle g_{2}(n)=\frac{A}{n}}$} & {$Np-A$} & {$1$} & {$A\geq A_{0}$} \tabularnewline
\hline 
\parbox[c]{2cm}{\centering ${\displaystyle g_{3}(n)=\frac{B}{\sqrt{n}}}$} & {$Np-B\sqrt{Np}$} & {${\displaystyle 1-\frac{B}{2\sqrt{Np}}}$} & {${\displaystyle B\geq\frac{A_{0}}{\sqrt{Np}}}$} \tabularnewline
\hline 
\end{tabular}
}%
}
\par\end{centering}
\caption{For each strategy $i \in \{1, 2, 3\}$, this table shows the computed values of $\langle L \rangle$ and $\sigma_L$, along with the reformulated version of Eq.~(\ref{condition3}) adapted to each specific function $g_i(\cdot)$.}
\label{tab3}
\end{table}

For the case $i = 3$, we needed to compute the expected value of the square root of a Gaussian random variable $X \sim \mathcal{N}(\xi, \sigma)$. To do so, we used an explicit formula from \cite{winkelbauer2012moments}, which confirmed that this value is approximately $\sqrt{\xi}$ in our particular setting.
Imposing the constraint $g_i(n) \leq 1/2$ and evaluating Eq.~(\ref{condition2}), we are able to provide lower bounds $N_{F}^{(i)}$ such that all the introduced conditions hold for $N \geq N_{F}^{(i)}$. The procedure used to obtain these results, as well as the justification for choosing $A = A_0$ and $B = A_0/\sqrt{n_{\mathrm{lim}}}$, is detailed in Appendix~\ref{appC}. The results are
\begin{equation}\label{res1}
\resizebox{\linewidth}{!}{$
\displaystyle
N_F^{(1)} = \max\left[ \frac{A_{0}}{g \cdot p},\; \frac{C_{F}^{2}}{4p} \left( \sqrt{1 - p} + \sqrt{(1 - p) + \frac{4l_{F}}{C_{F}^{2}(1 - g)}} \right)^2 \right]
$}
\end{equation}
\begin{equation}\label{res2}
\resizebox{\linewidth}{!}{$
\displaystyle
N_F^{(2)} = \max\left[ \frac{2A_{0}}{p},\; \frac{C_{F}^{2}}{4p} \left( \sqrt{1 - p} + \sqrt{(1 - p) + \frac{4(A_{0} + l_{F})}{C_{F}^{2}}} \right)^2 \right]
$}
\end{equation}
\begin{equation}\label{res3}
    N_F^{(3)}=\frac{1}{p}\max\left[\frac{4A_{0}^2}{n_{lim}}, n_{lim}\right],
\end{equation}
where $n_{lim}$ is the highest solution for $x$ of the equation $x^{3/2}-C_{F}\sqrt{1-p}\cdot x-\left(l_{F}+A_{0}\right)\sqrt{x}+\frac{A_{0}C_{F}}{2}\sqrt{1-p}=0$, which will be solved numerically.

\begin{figure}[!b]
    \centering
    \includegraphics[width=1\columnwidth]{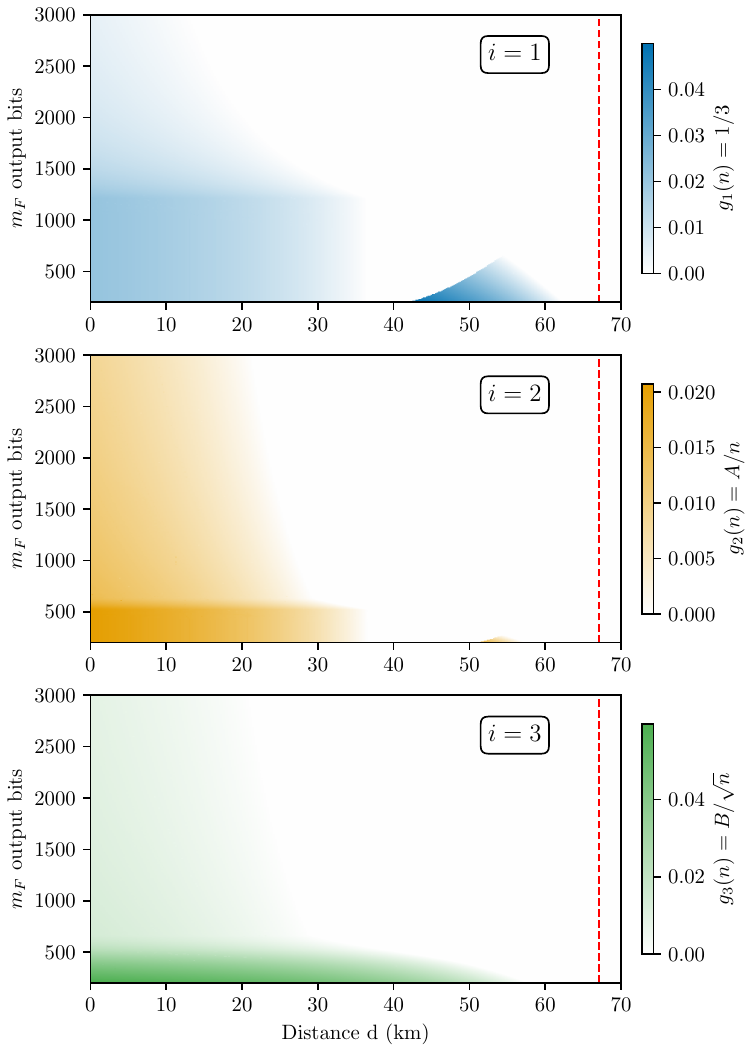}
    \caption{Optimal extra noise $P_{\mathrm{extra}}^{(\mathrm{opt},i)}$ as a function of link distance $d$ and output key length $m_F$. The three subplots correspond to each strategy, with $i = 1$ shown at the top, $i = 2$ in the middle, and $i = 3$ at the bottom. The values are obtained by numerically solving Eq.~(\ref{P_extra_opt}) using the parameters in Tab.~\ref{T1}. The red dashed line indicates the limit distance $d_{\mathrm{lim}}$ associated with the security threshold $Q_t=0.091$. }
    \label{fig:Pextra}
\end{figure}

Having established how to choose the initial number of photons to meet key length requirements, we now further elaborate on the motivation behind the controlled randomization phase. Without this step, the condition in Eq.~(\ref{condition3}) would cause $N_F \to \infty$ as $P_{\mathrm{flip}} \to 0$. To prevent this, we introduce an artificial noise term, $P_{\mathrm{extra}}$, applied to the raw keys. For the method to remain effective, $P_{\mathrm{extra}}$ should approach zero near the security threshold, ensuring $\hat{P}_{\mathrm{flip}} = P_{\mathrm{flip}}$, so the parameter estimation phase yields a reliable estimate of the QBER corresponding to the quantum channel in such critical scenarios.

The optimal value of $P_{\mathrm{extra}}$ for each parameter estimation strategy, denoted $P_{\mathrm{extra}}^{(\mathrm{opt}, i)}$, is determined according to the channel’s characteristics. Specifically, it minimizes the required initial number of single-photon pulses, $N_F^{(i)}$, and thus maximizes key distribution efficiency:
\begin{equation}\label{P_extra_opt}
\displaystyle
\resizebox{\linewidth}{!}{$
\begin{array}{l}
    P_{extra}^{(opt,i)}= \\
    \hfill \underset{P_{extra}}{\mathrm{argmin}}\left[N_{F}^{(i)}(P);\quad P=P_{extra}+P_{flip}-2P_{extra}P_{flip}\right]
\end{array}
$}
\end{equation}

In Fig.~\ref{fig:Pextra}, we show how the optimal extra noise, $P_{\mathrm{extra}}^{(\mathrm{opt},i)}$, depends on the link distance $d$ and the output key length requirement $m_F$. The plot shows that the added noise remains bounded within the considered range and decreases to zero as the distance approaches the limit $d_{\mathrm{lim}}$, which corresponds to the security threshold $Q_t=0.091$.

\section{Simulation Results}\label{5}
\begin{figure}
    \centering
    \includegraphics[width=1\columnwidth]{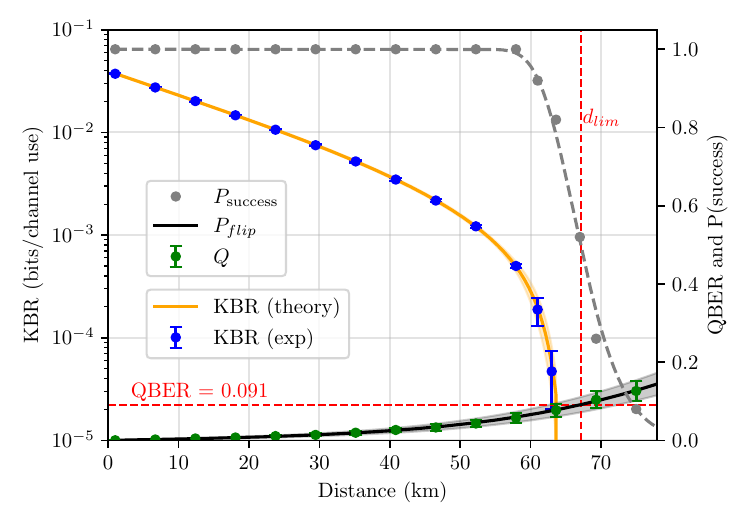}
    \caption{Behavior of our simulation setup compared with expected mean values (lines) and standard deviations (dashed areas). The simulation has been executed considering the parameters in Tab~\ref{T1}, selecting strategy $i=1$ ($g=1/3$) for parameter estimation, $N=2\cdot10^5$ as the number of initial photons and $P_{extra}=0$ as the artificial flip probability. Each experimental point and its error bar has been obtained running our BB84 variant $N_{iters}=50$ times. To validate the simulation results with the theoretical framework explained throughout this paper, we have used the second row of Tab.~\ref{tab3} and Eqs.~(\ref{Pflip}) and (\ref{Q}) to calculate the expected value and standard deviation of the QBER (right axis), Eq.~(\ref{KBR}) to estimate the $\mathrm{KBR}$ (left axis), and Eq.~(\ref{Psuccess}) for the protocol success probability $P(Q\leq 0.091)$ (right axis). As red dashed lines, we also plot the boundary $Q_t=0.091$ (horizontal) and the limit distance $d_{lim}$ (vertical) corresponding to such threshold. This is also the distance for which $\mathrm{KBR}_{\infty}=\lim _{N\rightarrow\infty}\mathrm{KBR}$ goes to 0.}
    \label{fig:BB84_plot}
\end{figure}

In this section, we present the final results on the performance of our variable-length version of BB84, introduced in Sect.~\ref{3}, and the proposed strategies for meeting predefined output length requirements, discussed in Sect.~\ref{4}. To do so, we first define the parameters typically used to evaluate this behavior.
One such parameter is the protocol success probability $P_{success}$, defined as the probability that the protocol does not abort after parameter estimation. In our case, we estimate this quantity using Eq.~(\ref{Q}) as
\begin{equation}\label{Psuccess}
    P_{success}=P(Q\leq Q_{t})=\Phi \left(\frac{Q_{t}-P_{flip}}{\sigma _Q}\right),
\end{equation}
where $Q_{t}=0.091$ and $ \Phi(x) = \frac{1}{\sqrt{2\pi}} \int_{-\infty}^{x} e^{-t^2/2} \,dt $ is the CDF of the standard normal distribution.
Another relevant parameter is the key bit rate (KBR), the number of secret output key bits generated per use of the quantum channel. We define it as:
\begin{equation}\label{KBR}
    \mathrm{KBR}=
\begin{cases} 
    M/N, & \text{with probability } P_{\text{success}} \\
    0, & \text{with probability } 1 - P_{\text{success}}
\end{cases}
\end{equation}
We estimate its average value and standard deviation as:
\begin{equation}\label{KBR_mean}
    \left\langle \mathrm{KBR}\right\rangle =P_{success}\left\langle M\right\rangle / N
\end{equation}
\begin{equation}\label{KBR_std}
\resizebox{\linewidth}{!}{$
    \sigma_{\mathrm{KBR}}=\frac{1}{N}\sqrt{P_{success}\sigma_{M}^{2}+P_{success}\left(1-P_{success}\right)\left\langle M\right\rangle ^{2}}
    $}
\end{equation}
These values can be calculated considering Tab.~\ref{tab3} to solve Eq.~(\ref{ml}) numerically and find $\left\langle M\right\rangle $, and to calculate $\sigma _M$ taking Eq.~(\ref{deltam}).
In Fig.~\ref{fig:BB84_plot}, we represent the performance of our protocol variant, presented in Sect.~\ref{3}. To simulate it, we use \texttt{NetSquid} for the quantum phase, while the post-processing is handled using third-party libraries \cite{cascade, foreman2025cryptomite}. We plot the simulation results for the $\mathrm{KBR}$ and $Q$ in terms of the link distance $d$ and compare them with the theoretical expectations. 
The resulting plot shows good agreement between simulation and theory. It reveals that the KBR is heavily limited by channel attenuation and drops to zero as $Q$ approaches the threshold $Q_t$. It also shows that the effective maximum distance is shorter than the asymptotic limit $d_{lim}$ (vertical red dashed line), due to finite key size effects.

\begin{figure}[!t]
    \centering
    \includegraphics[width=1\columnwidth]{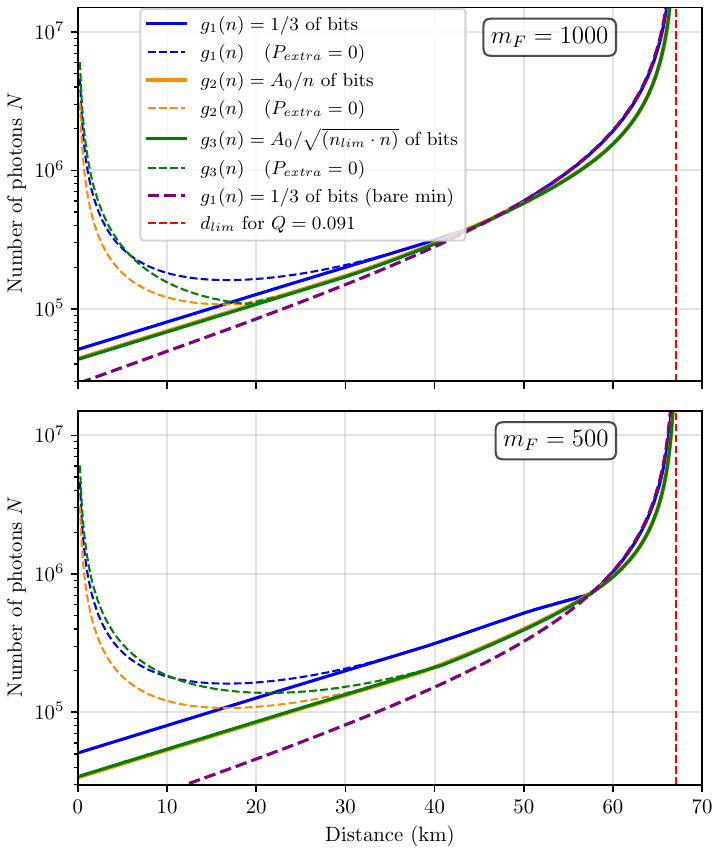}
    \caption{Representation of $N_F^{(i)}(d)$ for two output length specifications, $m_F = 1000$ bits (top) and $m_F = 500$ bits (bottom). They have been calculated using Eqs.~(\ref{res1}), (\ref{res2}) and (\ref{res3}) for each parameter estimation strategy in Tab.~\ref{tab3}. Each one of these three series also show the calculation corresponding to the case of $P_{extra} = 0$, where Bob does not add extra noise to the raw key. The violet dashed line shows the bare minimum number of photons, computed for $i=1$ and $P_{extra} = 0$ without considering restriction (\ref{condition3}). }
    \label{fig:N_F}
\end{figure}

Having obtained a benchmark for the performance of our simulation setup and compared it with the theoretically expected behavior, we will now test our reverse analysis approach. Fixing a value $m_F$ for the required output key length, we use Eqs.~(\ref{res1}), (\ref{res2}) and (\ref{res3}) to calculate the minimum number of photons needed to obtain such value for each parameter estimation strategy $g_i(\cdot)$ and in terms of the distance, obtaining $N_F^{(i)}(d,m_F)$. These quantities are represented in  Fig.~\ref{fig:N_F} for $m_F=1000$ and $m_F = 500$ bits required as output.
\begin{figure}
    \centering
    \includegraphics[width=1\columnwidth]{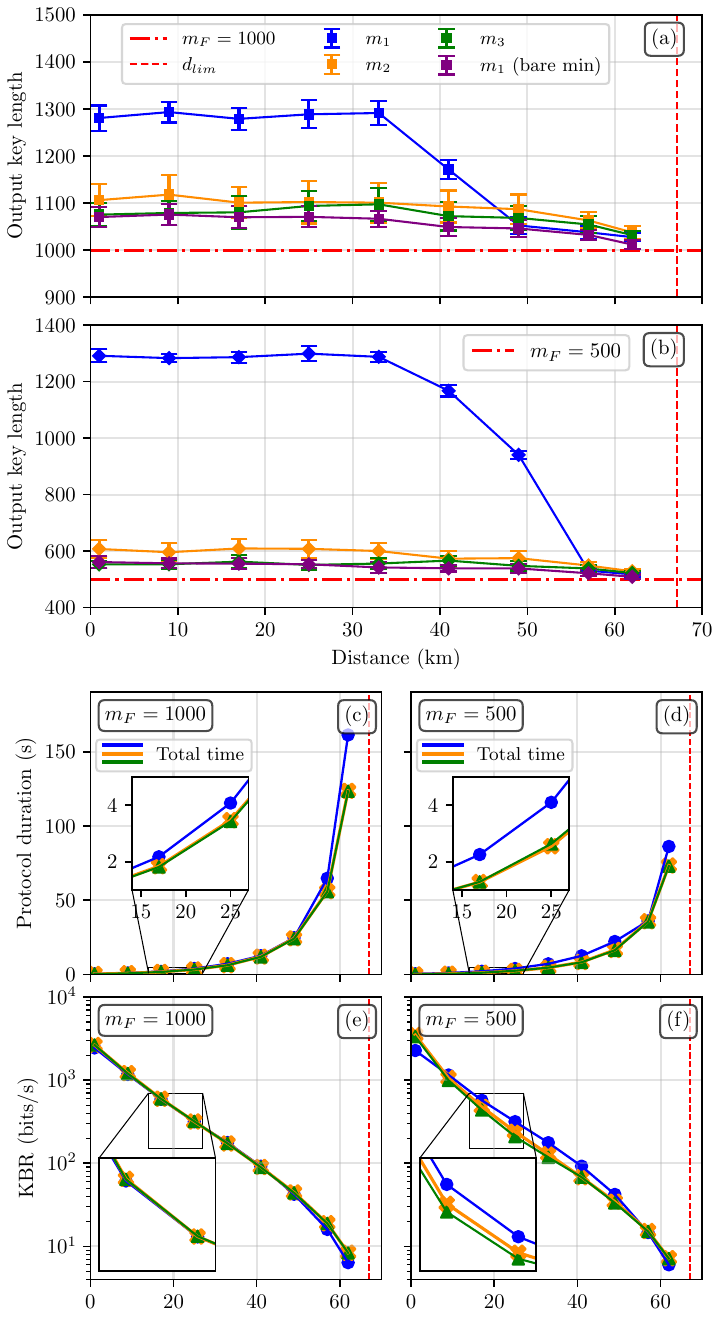}
    \caption{
    Performance of our reverse analysis, fixing $m_F$ as the required output key length and simulating the QKD protocol using $N = N_F^{(i)}(d, m_F)$. Output key length results are shown in subfigures~(a) and~(b) for $m_F = 1000$ and $m_F = 500$, respectively. The violet series represents the case of the bare minimum number of photons, computed for strategy $i=1$ without applying restriction~(\ref{condition3}).
    For each link distance $d$, we perform $N_{\text{iters}} = 20$ protocol simulations to obtain the mean output and its standard deviation, represented as error bars. Subfigures~(c) and~(d) show the simulated protocol duration, accounting for both the quantum transmission and classical post-processing stages. Subfigures~(e) and~(f) present the corresponding KBR in bits/s.
    In all subplots, the limit distance $d_{\text{lim}}$ is marked as a vertical red dashed line. The three parameter estimation strategies are clearly distinguished using the same color scheme as in Fig.~\ref{fig:N_F}.
    All simulations were carried out using the parameters listed in Table~\ref{T1}.
    }
    \label{fig:final_results}
\end{figure}
Using them for initialization, we will run the full QKD protocol to represent the output key length $m_i$ for $i=1,2,3$ and test if the inequality $m_i\geq m_F$ holds with the desired level of confidence. 
The results are shown in Fig.~\ref{fig:final_results}, which supports our main premise: a single execution of the protocol must yield at least $m_F$ secret key bits, as specified by the user, so that re-execution is not needed. We observe that all three strategies behave similarly, producing at least $m_F$ bits in most cases. Strategy $i=1$ tends to overperform at short distances, where our analysis is prioritizing an accurate estimation of the QBER using Eq.~(\ref{condition3}). On the other hand, strategies $i=2$ and $i=3$ align more closely with the output length requirement across both short and long distances, although they yield slightly lower KBR values. We also observe that they are marginally more efficient than strategy $i=1$ as the link distance approaches $d = d_{\mathrm{lim}}$.
A more detailed analysis shows that strategy $i=2$ slightly outperforms strategy $i=3$ in terms of KBR, while still meeting the output length requirement. This is because strategy $i=2$ uses a fixed number of bits for parameter estimation, chosen to be the minimum needed for an accurate estimation of the QBER.

\section{Conclusions}\label{6}

In this work, we have presented a variable-length adaptation of the BB84 protocol, specifically designed to generate secret keys of a target length under non-ideal conditions. 
Efficiency is a central design objective in this project, as our goal is to maximize the number of secure bits extracted per transmitted quantum signal while minimizing resource overhead during post-processing. This includes adjusting the fraction of discarded bits in parameter estimation, limiting communication and computation costs in error correction, and keeping the final key length as close as possible to the user-defined target, all while rigorously adhering to security requirements that guarantee the secrecy of the output key.
Our approach enables a dynamic configuration of the initial parameters of the protocol execution based on user-defined output constraints. To support both evaluation and reproducibility, we have developed Python-based simulation tools capable of modeling general QKD networks and protocols. They have been made publicly available on GitHub \cite{Martin_Megino_Non-ideal-QKDNs_2025}. While these tools were used to implement and test our proposed strategies, they were designed with broader applicability in mind and thus have the potential to be used in other research contexts.
The results confirm that our proposed strategies successfully allow to meet any output key length requirement in a single execution. Although excess key material may be produced, it can be stored for future use, making our strategy both effective and resource-efficient.

In addition to the future directions already mentioned, another natural continuation would be to apply our approach to improve key management strategies in the KM layer. This could help improve and strengthen the connection between the physical layer and the classical infrastructure. We emphasize the importance of making this interface more flexible, to improve usability and make QKD more accessible for researchers both inside and outside the quantum communication community. In particular, moving away from closed, black-box systems toward open and configurable devices is crucial, just as it has been in the development of other technologies. This flexible and multidisciplinary approach is essential for the broader development and scalability of QKD, as it moves from experimental prototypes and research-focused commercial devices toward real-world deployment and impactful applications.

\section{Acknowledgements}
This work has been partially supported by the MadQuantum-CM project, funded by the Regional Government of Madrid, the Spanish Government through the Recovery, Transformation and Resilience Plan, and the European Union through the NextGeneration EU funds;  by the 6G-INSPIRE project, PID2022-137329OB-C42, funded by MCIN/AEI/10.13039/501100011033/; and by the EU Horizon Europe project Quantum Security Networks Partnership (QSNP), under grant 101114043.
\vspace{0pt}

\onecolumn
\bibliography{tfm_bibliography2}
\bibliographystyle{quantum}

\section*{Appendices}
\appendix
\numberwithin{equation}{section}            
\renewcommand{\theequation}{\thesection\arabic{equation}} 
\section{Theoretical expectation of the QBER}
\label{appA}

Imagine that Alice sends the photon $i$ through the channel, and Bob is ready to measure it, turning his detector on during a certain time interval. Assuming the photon is initially in a pure state $\rho_{i,0} = \left|\Psi_{i}\right\rangle \left\langle \Psi_{i} \right|$, where $\left|\Psi_{i}\right\rangle = H^{b_{A,i}}X^{k_{A,i}}\left|0\right\rangle$, let us write the density matrix representing the qubit sent by Alice from Bob’s point of view. In an ideal and noiseless scenario, it is given by
\begin{equation}
    \rho_{i,f} = \frac{1}{4} \sum_{s \in \left\{ 0, 1, +, - \right\}} \left|s\right\rangle \left\langle s\right|,
\end{equation}
since Bob has no knowledge of $k_{A,i}$ or $b_{A,i}$.
However, in realistic scenarios, photons may be depolarized or lost, and undesired signals may be recorded instead. Let us describe the quantum state of the photon after transmission and Bob's basis selection (applying $X$ or $H$ gates) from his perspective, as
\begin{equation}
    \rho_{i,f} = P(X)\left|X\right\rangle \left\langle X\right| + \frac{1}{4} \sum_{s \in \left\{ 0, 1, +, - \right\}} \rho_{s},
\end{equation}
where the states $s = 0, 1$ correspond to cases in which Bob decodes in the same basis as Alice encodes. Here, $P(X) = P_{loss}\left(1 - P_{DCR}\right)$ is the probability that no signal is detected during the time window (neither the photon nor a dark count), which is represented by the ket $\left|X\right\rangle$. $\rho_{s}$ represents the density matrix associated with the four equiprobable states that Alice could have sent, and it takes the form

\begin{equation}\label{rho_s}
\displaystyle
\rho_{s} =  \frac{1}{2} \left(1 - P_{loss}\right) P_{DCR} \cdot \frac{\mathbb{I}}{2}
\displaystyle
+ \left(1 - \frac{1}{2} P_{DCR}\right) \left(1 - P_{loss}\right) \cdot \left[ (1 - P_{depolar}) \left|s\right\rangle \left\langle s\right| + P_{depolar} \frac{\mathbb{I}}{2} \right]
\displaystyle
\hfill + P_{loss} P_{DCR} \cdot \frac{\mathbb{I}}{2}
\end{equation}

The first term in Eq.~(\ref{rho_s}) accounts for events where the photon arrives during the time window, but a dark count occurs and is detected first (hence the factor $1/2$). The second term represents the case where the original photon, possibly depolarized, is detected before any dark count occurs. Finally, the third term corresponds to the case where the photon is lost, and a dark count is detected instead.

We are only interested in the cases where a signal is detected and the measurement basis coincides with the encoding basis. Let us denote Alice’s state as $\left|\Psi_{i}\right\rangle = \left|S\right\rangle$ and its orthogonal counterpart as $\left|\bar{\Psi}_{i}\right\rangle = \left|\bar{S}\right\rangle$. The density matrix conditioned to these events becomes
\begin{equation}
\displaystyle
\begin{array}{l}
\displaystyle
 \rho_{i,f} = \frac{1}{1 - P(X)} \rho_{S} = \frac{1}{1 - P(X)} \left[\left[ \frac{1}{2} (1 - P_{loss}) P_{DCR} \cdot \frac{\mathbb{I}}{2} + P_{loss} P_{DCR} \cdot \frac{\mathbb{I}}{2} \right]\right. + \hfill \\[0.7em]
 \displaystyle
 +  \left. \left[ \left(1 - \frac{1}{2}P_{\text{DCR}}\right)(1 - P_{\text{loss}}) \left[ (1 - P_{\text{depolar}})|S\rangle \langle S| + P_{\text{depolar}} \frac{\mathbb{I}}{2} \right] \right]\right] \\[0.7em]
 \displaystyle
 = \frac{1}{1 - P(X)} \left[\left[ \left(1 - \frac{1}{2} P_{DCR}\right) (1 - P_{loss}) (1 - P_{depolar}) \left|S\right\rangle \left\langle S\right| \right]\right. \\[0.7em]
 \displaystyle
 + \left. \left[ \left( \frac{1}{2} P_{DCR} (P_{loss} + 1) + P_{depolar} \left(1 - \frac{1}{2} P_{DCR}\right) (1 - P_{loss}) \right) \cdot \frac{\mathbb{I}}{2} \right] \right]  
\\[0.7em]
\displaystyle
= 2P_{flip} \cdot \frac{\mathbb{I}}{2} + \left(1 - 2P_{flip}\right) \left|S\right\rangle \left\langle S\right|
= P_{flip} \left( \left|S\right\rangle \left\langle S\right| + \left|\bar{S}\right\rangle \left\langle \bar{S} \right| \right) + \left(1 - 2P_{flip}\right) \left|S\right\rangle \left\langle S\right|  = \\[0.7em]
\displaystyle
\hfill \left(1 - P_{flip}\right) \left|S\right\rangle \left\langle S\right| + P_{flip} \left|\bar{S}\right\rangle \left\langle \bar{S} \right|,
\end{array}
\end{equation}
where
\begin{equation}
\displaystyle
    P_{flip} = \frac{ \frac{1}{4} P_{DCR} (P_{loss} + 1) + \frac{P_{depolar}}{2} \left(1 - \frac{1}{2} P_{DCR}\right) (1 - P_{loss}) }{ 1 - P_{loss} (1 - P_{DCR}) }.
\end{equation}

\section{Error rate as a random variable}
\label{appB}

Starting from $N$ photons, we want to estimate the number of bits surviving channel transmission and key sifting. For a single bit, the probability of surviving both processes is $p$, defined in Eq.~(\ref{p}), assuming both measurement bases are equiprobable. Then, the number of bits surviving this phase follows a binomial distribution,
\begin{equation}
    Y \sim \mathrm{Bin}(N, p) \equiv Np \pm \sqrt{Np(1 - p)},
\end{equation}
where we use this notation to indicate the mean and standard deviation. It is useful to note that, in our scenario, $Np$ is large enough to approximate the binomial distribution by a normal distribution:
\begin{equation}
    \mathrm{Bin}(N, p) \simeq \mathcal{N}(Np, Np(1 - p)).
\end{equation}

Being $Y$ the number of key bits at this point, we will use a fraction  $g(Y) \leq \frac{1}{2}$ of them to estimate the QBER. The number of remaining bits is then described by the random variable 
\begin{equation}
    L \sim (1 - g(Y)) \cdot Y.
\end{equation}

To estimate the QBER, we consider two bit strings of length $K = g(Y) \cdot Y$. The bitwise probability of a mismatch between Alice and Bob's raw keys is denoted by $\hat{P}_{\mathrm{flip}}$, defined in Eq.~(\ref{P_flip_hat}) as the effective bit flip probability. It accounts for the combined effect of quantum channel noise and the artificial randomization introduced by Bob. Let $Z$ be the total number of bit flips between the two strings. Then $Z$ follows a binomial distribution conditioned on the value of $K$,
\begin{equation}
    Z \mid (K=k) \sim \mathrm{Bin}(k, \hat{P}_{flip}).
\end{equation}

The ratio $\hat{Q} = Z/K$ serves as our estimator for the QBER. To find the distribution of $\hat{Q}$, we consider the joint distribution of $Z$ and $K$, and then marginalize over $K$.
Given $K = k$, the distribution of $Z$ is
\begin{equation}
P(Z = z \mid K = k) = \binom{k}{z} {\hat{P}_{flip}}^z (1 - \hat{P}_{flip})^{k - z}, \quad z = 0, \ldots, k.
\end{equation}
On the other hand, since $K = g(Y) \cdot Y$ and $Y \sim \mathrm{Bin}(N, p)$, the distribution of $K$ is given by
\begin{equation}
P(K = k) = \sum_{y\in\mathbb{N}: \;g(y)\cdot y = k} \binom{N}{y} p^y (1 - p)^{N - y}.
\end{equation}
Since $\hat{Q} = Z/K$ is a function of $Z$ and $K$, its distribution is obtained by marginalizing over $K$:
\begin{equation}
P(\hat{Q} = q) = \sum_k P\left(\frac{Z}{K} = q \mid K = k\right) \cdot P(K = k).
\end{equation}
Given that $Z \mid (K=k) \sim \mathrm{Bin}(k, \hat{P}_{flip})$, the conditional distribution of $Q \mid K$ is
\begin{equation}
\hat{Q} \mid (K=k) \sim \frac{\mathrm{Bin}(k, \hat{P}_{flip})}{k}.
\end{equation}
If $K$ is sufficiently large, we can approximate this as
\begin{equation}
\hat{Q} \mid (K=k) \simeq \mathcal{N}\left(\hat{P}_{flip}, \frac{\hat{P}_{flip}(1 - \hat{P}_{flip})}{k}\right).
\end{equation}
Finally, making a further approximation using the expected value of $K$, we get
\begin{equation}
\hat{Q} \simeq \mathcal{N}\left(\hat{P}_{flip}, \frac{\hat{P}_{flip}(1 - \hat{P}_{flip})}{\mathbb{E}[K]}\right).
\end{equation}

We can particularize this expression for the different options for $g_i(\cdot )$:

\begin{table}[ht]
\renewcommand*\arraystretch{2}
\centering
\begin{tabular}{|P{4cm}|P{3cm}|P{7cm}|}
\hline
$g_i(n)$ & $\mathbb{E}[Y\cdot g_i(Y)]$ & $Q$ \\
\hline
$ g_1(n)=g \in (0, 1/2)$ & $gNp$ & $\hat{Q}_1 = \hat{P}_{flip} \pm \sqrt{\frac{\hat{P}_{flip}(1 - \hat{P}_{flip})}{gNp}}$ \\
\hline
$g_2(n) = A/n $ & $A$ & $\hat{Q}_2 = \hat{P}_{flip} \pm \sqrt{\frac{\hat{P}_{flip}(1 - \hat{P}_{flip})}{A}}$ \\
\hline
$g_3(n) = B/\sqrt{n} $ & $B\sqrt{Np}$ & $\hat{Q}_{3}=\hat{P}_{flip}\pm\sqrt{\frac{\hat{P}_{flip}\left(1-\hat{P}_{flip}\right)}{B\sqrt{Np}}}$ \\
\hline
\end{tabular}
\caption{Parameter estimation strategies $g_i(n)$ for $i = 1, 2, 3$, with their corresponding expected sample sizes $\mathbb{E}[Y \cdot g_i(Y)]$ and error rate estimation uncertainty. These expressions are used to evaluate condition~(\ref{condition3}).}
\label{tab:example}
\end{table}

\section{Calculating $N_F^{(i)}$}
\label{appC}

Having derived the expression $\left\langle L \right\rangle - C_{F} \cdot \sigma_{L} \geq  l_{F}$ in Eq.~(\ref{condition2}), let us calculate $\left\langle L \right\rangle$ and $\sigma_{L}$. To do that, we remember that $L=(1-g(Y))\cdot Y$ where $Y \sim \mathrm{Bin}(N, p)$.
With these quantities, we are in position to give a first estimate of $N_F^{(i)}$. Let us address this problem for three versions of $g_i(n)$. Each defines a different lower bound on the number of photons needed for accurate parameter estimation, sufficient key bit supply, and a suitable fraction of bits allocated for parameter estimation.

\paragraph*{Case 1: $g_1(n) = g \in (0, 1)$.} 

This strategy consists of taking a fixed fraction of the sifted key bits for parameter estimation.
Given $Y \sim \mathrm{Bin}(N, p)$, we define the number of bits remaining in the key as $L \sim (1 - g)\cdot X$. Applying the Gaussian approximation, we find
\begin{equation}
    L \simeq \mathcal{N}\left((1 - g)\cdot Np, \, (1 - p)(1 - g)^2 \cdot Np\right),
\end{equation}
so that
\begin{equation}
L_1 = (1 - g) \left(Np \pm \sqrt{Np(1 - p)}\right).
\end{equation}
Including these values into Eq.~(\ref{condition2}) we get
\begin{equation}\label{cond1app}
    N^{(1)}\geq \frac{C_F^2}{4p}\cdot \left[ \sqrt{1-p} + \sqrt{1-p+\frac{4l_F}{C_F^2\cdot (1-g)}} \right]^2.
\end{equation}
To finally obtain $N_F^{(i)}$, we also need to assure that the QBER is properly estimated, according to Eq.~(\ref{condition3}). From that equation, we defined
\begin{equation}
    A_0:=\frac{1}{\Gamma(\varepsilon)^2}\left(\frac{1}{\hat{P}_{flip}}-1\right).
\end{equation}
On the one hand, we require $N^{(1)}$ to verify the condition already stated in Eq.~(\ref{cond1app}). On the other hand, since $g \leq 1/2$ and we need $Np \geq A_{0}/g$ for accurate error rate estimation, we conclude:
\begin{equation}
N_{F}^{(1)} = \frac{1}{p} \max \left[ \frac{A_{0}}{g}, \, \frac{C_F^2}{4}\left( \sqrt{1 - p} + \sqrt{(1 - p) + \frac{4l_{F}}{C_F^2(1 - g)}} \right)^2 \right].
\end{equation}

\paragraph*{Case 2: $g_2(n) = A/n$ for $n \geq 2A$.}
This second strategy takes a fixed number $A$ of bits from the sifted key for parameter estimation, ensuring that the total key length is sufficient to allow it.
In this case, we subtract a constant $A$ from the sifted key, assuming $n \simeq Np \geq 2A$. Then $L_2 \sim Y - A$ and we find
\begin{equation}
L_2 = (Np - A) \pm \sqrt{Np(1 - p)}
\end{equation}
Including these values into Eq.~(\ref{condition2}) we get
\begin{equation}\label{cond2app}
    N^{(2)}\geq \frac{C_F^2}{4p}\cdot \left[ \sqrt{1-p} + \sqrt{1-p+\frac{4(l_F + A)}{C_F^2}} \right]^2.
\end{equation}
To satisfy the constraint $Np \cdot g(Np) =A\geq A_{0}$ for correct QBER estimation, the best choice is $A = A_{0}$. Then, we require that
\begin{equation}
\resizebox{\linewidth}{!}{$
\displaystyle
\left\langle L \right\rangle - C_{F} \cdot \sigma_{L} \geq l_{F}
\Longleftrightarrow Np - A_{0} - C_{q} \sqrt{Np(1 - p)} \geq l_{F}
\Longleftrightarrow Np \geq \frac{1}{4}\left[ C_{F} \sqrt{1 - p} + \sqrt{C_{F}^{2}(1 - p) + 4(A_{0} + l_{F})} \right]^2.
$}
\end{equation}
Additionally, $g(Np) \leq 1/2$ implies $Np \geq 2A$, so we finally obtain
\begin{equation}
\displaystyle
N_{F}^{(2)} =
\displaystyle
\frac{1}{p} \max \left[ 2A_{0}, \, \frac{C_F^2}{4}\cdot \left[ \sqrt{1-p} + \sqrt{1-p+\frac{4(l_F + A)}{C_F^2}} \right]^2 \right].
\end{equation}

\paragraph*{Case 3: $g_3(n) = B/\sqrt{n}$ for $n \geq 4B$.}

This third strategy uses a number of bits for parameter estimation that scales with $\sqrt{n}$. It was suggested to be optimal by Scarani and Renner in \cite{scarani2008quantum} and its behavior lies between Case 1 and Case 2. We will show that it is nearly optimal, although Case 2 proves more suitable for our purposes.
We start with the expression
\begin{equation}
L = \left(1 - \frac{B}{\sqrt{Y}}\right)Y = Y - B\sqrt{Y}.
\end{equation}
Since this expression is nonlinear, a direct normal approximation is not straightforward. However, using a first-order Taylor expansion, we estimate the expected value and variance of $L$ \cite{winkelbauer2012moments} as
\begin{equation}
\mathbb{E}[L] \simeq \mathbb{E}[Y] - B\sqrt{\mathbb{E}[Y]},
\displaystyle
\quad \mathrm{Var}[L] \simeq \left(1 - \frac{B}{2\sqrt{\mathbb{E}[Y]}}\right)^2 \mathrm{Var}[Y].
\end{equation}
Then, using the Gaussian approximation we get
\begin{equation}
L_3 \simeq \left(1 - \frac{B}{\sqrt{Np}}\right)Np \pm \left(1 - \frac{B}{2\sqrt{Np}}\right)\sqrt{Np(1 - p)}.
\end{equation}
Including these values into Eq.~(\ref{condition2}) we find:
\begin{equation}\label{cond3app}
\displaystyle
N^{(3)} \geq \frac{1}{4p} \cdot \left[ B + C_F \sqrt{1 - p} + \sqrt{\left( B + C_F \sqrt{1 - p} \right)^2 + 4 \left( l_F - \frac{C_F B}{2} \sqrt{1 - p} \right)} \right]^2.
\end{equation}
In this last case, accounting for accuracy in the estimation of the QBER according to Eq.~(\ref{condition3}) is slightly more complicated. First, we find an initial bound for $Np$:
\begin{equation}
\begin{aligned}
Np \cdot g(Np) \geq A_{0} &\Longrightarrow B \geq \frac{A_{0}}{\sqrt{Np}} \\
\left\langle L \right\rangle - C_{F} \cdot \sigma_{L} \geq l_{F} &\Longrightarrow B \leq \frac{Np - C_{F} \sqrt{Np(1 - p)} - l_{F}}{\sqrt{Np} - \frac{C_{F}}{2} \sqrt{1 - p}}.
\end{aligned}
\end{equation}
Combining both inequalities, we get:
\begin{equation}
\displaystyle
\frac{A_{0}}{\sqrt{Np}} \leq \frac{Np - C_{F} \sqrt{1 - p} \sqrt{Np} - l_{F}}{\sqrt{Np} - \frac{C_{F}}{2} \sqrt{1 - p}} \Longrightarrow 
(Np)^{3/2} - C_{F} \sqrt{1 - p} \cdot Np - (l_{F} + A_{0}) \sqrt{Np}  + \frac{A_{0} C_{F}}{2} \sqrt{1 - p} \geq 0.
\end{equation}
Let $n_{lim}$ be the smallest solution to this inequality. Then $Np \geq n_{lim}$ is a sufficient condition. We can take
\begin{equation}
B = \frac{A_{0}}{\sqrt{n_{lim}}}.
\end{equation}
Also, since $g(Np) \leq 1/2$ implies $Np \geq 4A_{0}^2/n_{lim}$, we finally impose
\begin{equation}
\displaystyle
\begin{array}{l}
\left\langle L \right\rangle - C_{F} \cdot \sigma_{L} \geq l_{F}
\Longleftrightarrow
Np - (B + C_{F} \sqrt{1 - p}) \sqrt{Np} + \frac{BC_{F}}{2} \sqrt{1 - p} - l_{F} \geq 0 \Longleftrightarrow\\[0.5em]
 Np \cdot \sqrt{n_{lim}} - C_{F} \sqrt{1 - p} \cdot \sqrt{Np \cdot n_{lim}} - (l_{F} + A_{0}) \sqrt{n_{lim}} + \frac{A_{0} C_{F}}{2} \sqrt{1 - p} \geq 0.
\end{array}
\end{equation}
As expected, the solution is consistent with $n_{lim}$. Thus:
\begin{equation}
N_{F}^{(3)} = \frac{1}{p} \max \left[ n_{lim}, \, \frac{4A_{0}^2}{n_{lim}} \right]
\end{equation}
 
\end{document}